\documentclass[12pt]{article}
\usepackage{epsfig}
\usepackage{amsmath}
\usepackage{hhline}
\usepackage{amssymb}
\usepackage{times}
\usepackage{cite}

\newlength{\dinwidth}
\newlength{\dinmargin}
\begin{document}  
% The rest
\newcommand{\pom}{{I\!\!P}}
\newcommand{\reg}{{I\!\!R}}
\def\gsim{\,\lower.25ex\hbox{$\scriptstyle\sim$}\kern-1.30ex%
\raise 0.55ex\hbox{$\scriptstyle >$}\,}
\def\lsim{\,\lower.25ex\hbox{$\scriptstyle\sim$}\kern-1.30ex%
\raise 0.55ex\hbox{$\scriptstyle <$}\,}
\newcommand{\trm}{m_{\perp}}
\newcommand{\trp}{p_{\perp}}
\newcommand{\trmm}{m_{\perp}^2}
\newcommand{\trpp}{p_{\perp}^2}
\newcommand{\alp}{\alpha_s}
\newcommand{\alps}{\alpha_s}
\newcommand{\sqrts}{$\sqrt{s}$}
\newcommand{\LO}{$O(\alpha_s^0)$}
\newcommand{\Oa}{$O(\alpha_s)$}
\newcommand{\Oaa}{$O(\alpha_s^2)$}
\newcommand{\PT}{p_{\perp}}
\newcommand{\JPSI}{J/\psi}
\newcommand{\PO}{I\!\!P}
\newcommand{\xbj}{x}
\newcommand{\xpom}{x_{\PO}}
\newcommand{\dgr}{^\circ}
\newcommand{\gev}{\,\mbox{GeV}}
\newcommand{\GeV}{\rm GeV}
\newcommand{\xp}{x_p}
\newcommand{\xpi}{x_\pi}
\newcommand{\xg}{x_\gamma}
\newcommand{\xgj}{x_\gamma^{jet}}
\newcommand{\xpj}{x_p^{jet}}
\newcommand{\xpij}{x_\pi^{jet}}
\renewcommand{\deg}{^\circ}
\newcommand{\qsq}{\ensuremath{Q^2} }
\newcommand{\gevsq}{\ensuremath{\mathrm{GeV}^2} }
\newcommand{\et}{\ensuremath{E_t^*} }
\newcommand{\rap}{\ensuremath{\eta^*} }
\newcommand{\gp}{\ensuremath{\gamma^*}p }
\newcommand{\dsiget}{\ensuremath{{\rm d}\sigma_{ep}/{\rm d}E_t^*} }
\newcommand{\dsigrap}{\ensuremath{{\rm d}\sigma_{ep}/{\rm d}\eta^*} }
% Journal macro
\def\Journal#1#2#3#4{{#1} {\bf #2}, #4 (#3)}
\def\NCA{\em Nuovo Cimento}
\def\NIM{\em Nucl. Instrum. Methods}
\def\NIMA{{\em Nucl. Instrum. Methods} {\bf A}}
\def\NPB{{\em Nucl. Phys.}   {\bf B}}
\def\PLB{{\em Phys. Lett.}   {\bf B}}
\def\PRL{\em Phys. Rev. Lett.}
\def\PRD{{\em Phys. Rev.}    {\bf D}}
\def\PR{{\em Phys. Rev.}    }
\def\ZPC{{\em Z. Phys.}      {\bf C}}
\def\ZP{{\em Z. Phys.}      }
\def\EJC{{\em Eur. Phys. J.} {\bf C}}
\def\EJA{{\em Eur. Phys. J.} {\bf A}}
\def\CPC{\em Comp. Phys. Commun.}
\begin{titlepage}

\begin{flushleft}
DESY 04-247 \hfill ISSN 0418-9833 \\
December 2004
\end{flushleft}

\vspace{2cm}

\begin{center}
\begin{Large}

{\boldmath \bf      
    Measurement of Dijet Cross Sections in \boldmath$ep$ Interactions 
     with\\  a Leading  Neutron  at HERA
}

\vspace{2cm}

H1 Collaboration

\end{Large}
\end{center}

\vspace{2cm}

\begin{abstract}
\noindent
Measurements are reported of     the production of dijet events with a 
leading neutron in $ep$ interactions at HERA.
Differential cross sections for photoproduction 
and deep inelastic scattering  are presented as a function of 
several kinematic variables.
Leading order QCD simulation programs are compared with the measurements.
Models in which the real or virtual photon interacts with a parton 
of an exchanged pion are able to describe the data. Next-to-leading 
order perturbative QCD calculations based on pion exchange
are found to be in good agreement with the measured cross sections.
The fraction of leading neutron dijet events with respect to all
dijet events is also determined. The dijet events with a
leading neutron
have a lower fraction of resolved photon processes than do the
inclusive dijet data.
\end{abstract}

\vspace{1.5cm}

\begin{center}
Submitted to $Eur.~Phys.~J.~C$
\end{center}

\end{titlepage}

\begin{flushleft}
%-- H1AUTS Author list by names 
%-- Status: Tue Nov  2 12:19:12 MET 2004  Number of authors = 307 

A.~Aktas$^{10}$,               %DESY-ST        03/2            Aktas               
V.~Andreev$^{26}$,             %LPI -PD        8/88            Andreev             
T.~Anthonis$^{4}$,             %ANTW-ST        11/99           Anthonis            
S.~Aplin$^{10}$,               %DFLC-PD        01/04           Aplin               
A.~Asmone$^{34}$,              %ROME-ST        07/2            Asmone              
A.~Babaev$^{25}$,              %ITEP-PD        8/88            Babaev              
S.~Backovic$^{31}$,            %PODG-PD        03/2            Backovic            
J.~B\"ahr$^{39}$,              %ZEUT-PD        8/88            Baehr               
A.~Baghdasaryan$^{38}$,        %YERE-PD        09/03           Baghdasaryan        
P.~Baranov$^{26}$
% ...

%-- H1 Institutes 
\bigskip{\it
% ...
$ ^{4}$ Inter-University Institute for High Energies ULB-VUB, Brussels;
          Universiteit Antwerpen, Antwerpen; Belgium$^{ c}$ \\
% ...
$ ^{10}$ DESY, Hamburg, Germany \\
% ...
$ ^{25}$ Institute for Theoretical and Experimental Physics,
          Moscow, Russia$^{ l}$ \\
 $ ^{26}$ Lebedev Physical Institute, Moscow, Russia$^{ e}$ \\
% ...
$ ^{31}$ Faculty of Science, University of Montenegro,
          Podgorica, Serbia and Montenegro \\
% ...
$ ^{34}$ Dipartimento di Fisica Universit\`a di Roma Tre
          and INFN Roma~3, Roma, Italy \\
% ...
$ ^{38}$ Yerevan Physics Institute, Yerevan, Armenia \\
 $ ^{39}$ DESY, Zeuthen, Germany \\
% ...
$ ^c$ Supported by FNRS-FWO-Vlaanderen, IISN-IIKW and IWT
      and  by Interuniversity
Attraction Poles Programme,
      Belgian Science Policy \\
% ...
$ ^e$ Supported by the Deutsche Forschungsgemeinschaft \\
% ...
$ ^l$ Partially Supported by Russian Foundation
      for Basic Research, grant    no. 00-15-96584 \\
}
\end{flushleft}

\newpage

\section{Introduction}

Previous HERA measurements
\cite{H1LN,zeuslninc}  show that the cross section for the 
semi-inclusive $ep$ scattering process
 \begin{equation}  e+p\rightarrow e+n+X, \label{eq1}\end{equation}

\noindent
where the 
leading neutron carries more than 70\% of the proton beam 
energy, is reasonably well described by the 
pion exchange mechanism [3--7].
% %\cite{Sullivan-Przy}. 
In this picture, the virtual photon interacts 
with a parton from the pion. Constraints on the pion structure function 
are thus obtained.  However, the Soft Colour Interaction model,
in which colour neutral partonic subsystems are formed by non-perturbative
soft gluon exchanges \cite{LEPTO, sci}, describes the data equally 
well~\cite{H1LN}.

In the present analysis, the leading neutron production mechanism 
is investigated further by requiring that the system $X$ in (1) contains
 two jets with large transverse momenta
\begin{equation}  e+p\rightarrow e+n+jet+jet+X. \label{eq2}\end{equation}

\noindent
 This allows more detailed comparisons of the measurements to be made with
 model predictions.
% In such processes, more detailed comparisons of the measurements with
% model predictions can be performed, due to the presence of an
% exclusive hadronic final state. 
 In addition, the jet energy provides a hard scale which
allows the comparison of perturbative QCD with the data 
for all photon virtualities $Q^2$.
%, including the photoproduction regime.
The cross sections are measured in both photoproduction 
($Q^2<10^{-2}~\GeV^2$)  and deep inelastic scattering 
(DIS, $2<Q^2<80~\GeV^2$).
They are given as a function of $Q^2$ and of the kinematic variables of 
the jets. Monte Carlo predictions based on leading order (LO) QCD models
are compared to the data, as are next-to-leading order (NLO) QCD
calculations \cite{Klasen}.
Furthermore, a detailed comparison of dijet production with
and without the  requirement of a leading neutron is made.
In the photoproduction regime, similar studies
have been reported by the ZEUS Collaboration~\cite{ZEUSln}.

%%%%%%%%%%%%%%%%% Event Kinematics %%%%%%%%%%%%%%%%%%%%%%%%%%%%%%%%%

\section{Event Kinematics and Reconstruction}

\begin{figure}[h]
\epsfig{file=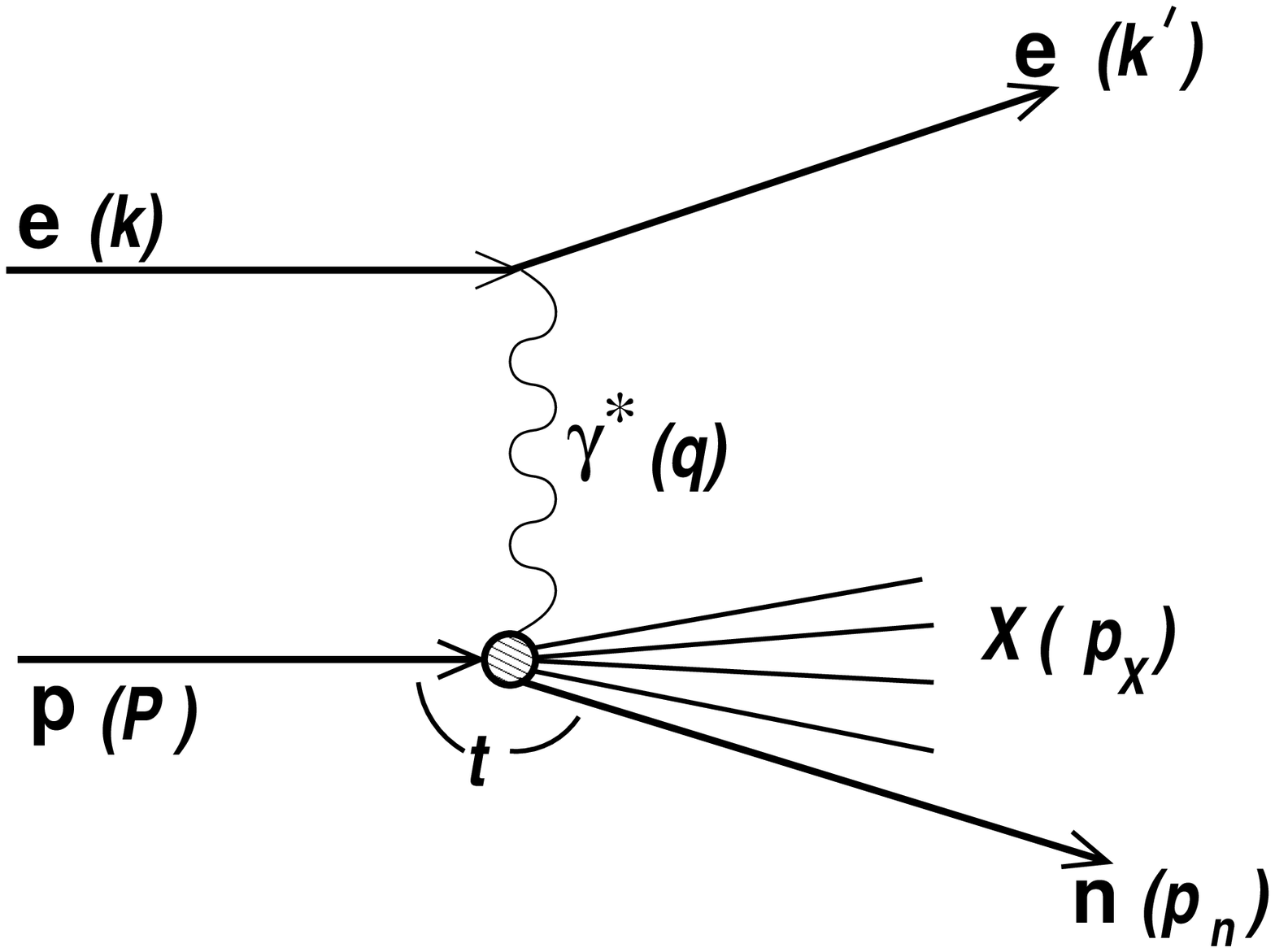,height=60mm}
\hspace*{5mm}
\epsfig{file=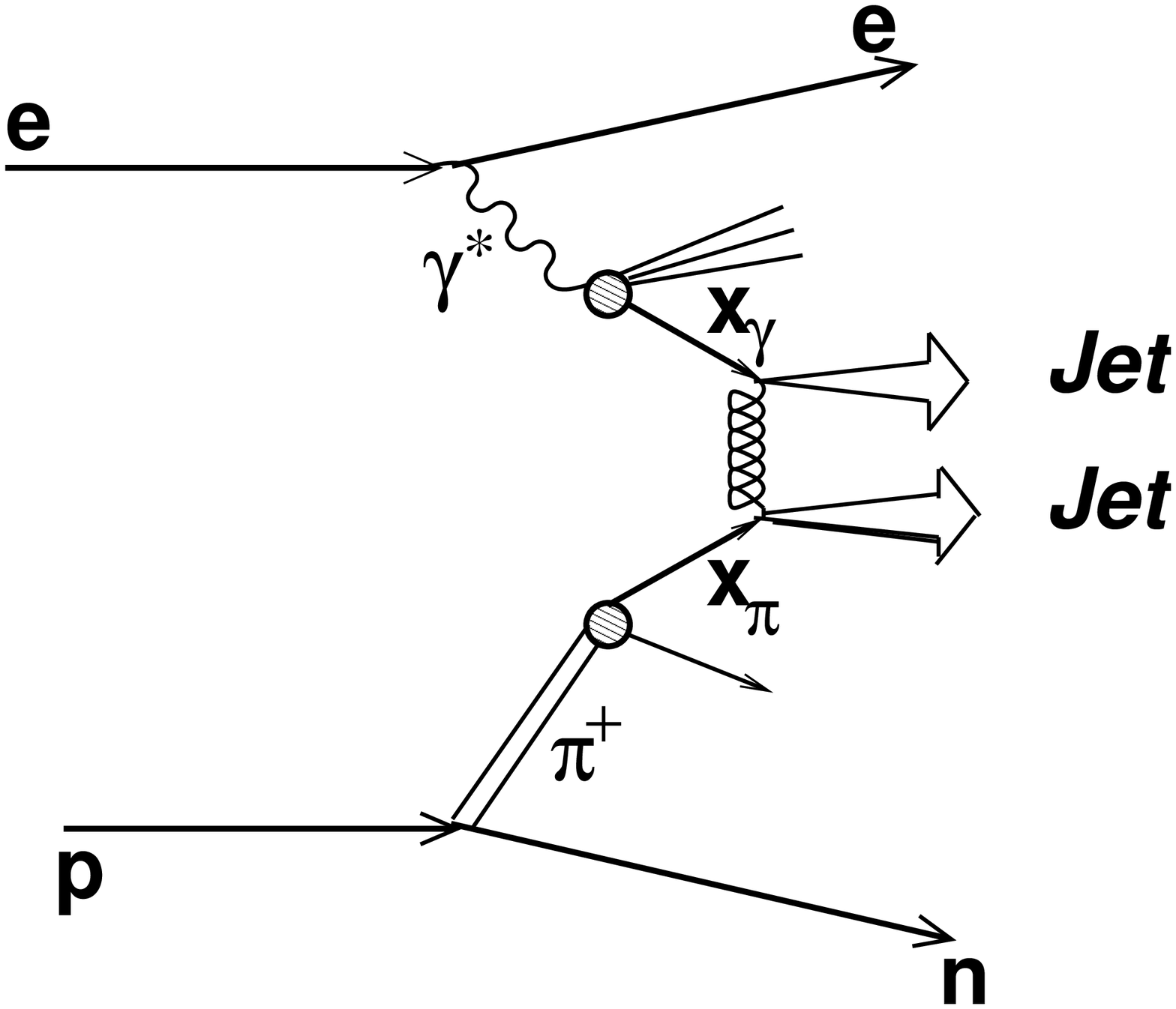,height=60mm}
\caption{ (a) A diagram for the process $e+p\rightarrow e+n+X$, 
(b) a diagram for the dijet production process $e+p\rightarrow e+n+jet+jet+X$ 
assuming  this proceeds via pion exchange.}

\vspace*{-73mm}
\hspace*{25mm}{\large \bf (a)} \hspace*{78mm} {\large \bf (b)} 

\vspace*{73mm}
\label{diagram1}
\end{figure}

The semi-inclusive reaction~(\ref{eq1}) is sketched
in Fig.~\ref{diagram1}a, in which the $4$-vectors of the incoming 
and outgoing particles and of the exchanged photon are indicated.
Figure~\ref{diagram1}b depicts the dijet production reaction (\ref{eq2}) 
under the assumption that it is mediated by pion exchange.

The standard Lorentz invariant kinematic variables used to describe
high energy $ep$ interactions are the centre-of-mass energy squared $s$,
the four-momentum transfer squared $Q^2$ and the inelasticity $y$:
\begin{eqnarray}
  s  &\equiv& (k+P)^2=4E_eE_p, \nonumber\\
 Q^2 &\equiv& -q^2= -(k-k')^2=4E_eE_e'\cos ^2\left(\frac{\theta '_e}{2}\right),\\
 y &\equiv& \frac{(q\cdot P)}{(k\cdot P)}=
 1-\frac{E_e'}{E_e} \sin ^2\left(\frac{\theta '_e}{2}\right).\nonumber
\end{eqnarray}

\noindent
These are determined from the energies $E_e$ and $E_p$ of the 
 lepton and proton beams, respectively, and from the energy $E_e'$ and
 polar angle $\theta_e'$ of the scattered 
lepton in the laboratory frame\footnote{The right-handed H1 coordinate 
system has its positive  $z$ direction along the proton
   beam direction and its origin at the nominal interaction point.}.
\vspace*{2mm}

%%%%%%%%%%%%%%%%%%%%% FIG.1 %%%%%%%%%%%%%%%%%%%%%%%%%%%%%%%%%%%%%%%%%%

\noindent
Two more invariant variables, $x_L$ and $t$, are used to
describe  the  kinematics of the semi-inclusive reaction~(\ref{eq1}):
\begin{eqnarray}
 x_L &\equiv& \frac{(q\cdot p_n)}{(q\cdot P)}\simeq \frac{E_n}{E_p}
  \nonumber\\ \hspace*{-18mm}
 \textrm{and} \hspace*{4mm}t  &\equiv& (P-p_n)^2\simeq 
-\frac{p_{Tn}^2}{x_L}-(1-x_L)\left(\frac{m_n^2}{x_L}-m_p^2\right),
\end{eqnarray}

\noindent
where $E_n$ is the neutron energy, $p_{Tn}$ is the 
momentum component of the neutron
transverse  to the direction of the incident proton and $m_n$ and $m_p$ 
are the neutron and proton masses, respectively. 
Experimentally, $x_L$ and $t$ are
determined from the measured energy and scattering angle of the 
leading  neutron.

In the pion exchange model, the photon interacts with a pion emitted from the
proton. In this model, process (2) is represented by diagrams as sketched in  
Fig.~\ref{diagram1}b. The quantity $\xpi$ denotes, neglecting masses,
the fraction of the 4-momentum of the pion participating 
in the hard interaction\footnote{The definitions of the variables $x_L$ and $x_\pi$ 
        are similar to the
        definitions of the variables $(1-\xpom)$ and $\beta$, used in H1 
        analyses of diffractive processes~\cite{f2d}.}.
It is related to $x_p$, the fraction of the 4-momentum of the  proton which 
enters the hard interaction, via $x_p=x_\pi(1-x_L)$.

The quantity $\xg$ is the fraction of the 4-momentum of the photon
which participates in the hard interaction.
If the virtual photon is ``resolved'' and participates 
in the hard interaction via its partonic content,
then $\xg<1$. If the interactions are ``direct'', i.e. the entire
photon enters the hard scattering process, then $\xg=1$.

The quantities $\xgj$, $\xpij$ and $x_p^{jet}$ , which are estimators for
$\xg$, $\xpi$ and $\xp$, can be defined in dijet events using  the 
 jet transverse energies $E_T^{jet}$ and 
  pseudorapidities $\eta^{jet}$   according to:

\begin{eqnarray}
x_\gamma^{jet}&=&\frac{(E_T^{jet1}e^{-\eta^{jet1}}+
                 E_T^{jet2}e^{-\eta^{jet2}})}{2yE_e},~~~~
x_p^{jet}=\frac{(E_T^{jet1}e^{\eta^{jet2}}+
                 E_T^{jet2}e^{\eta^{jet2}})}{2E_p} \nonumber \\
\textrm{and} ~~~~
x_\pi^{jet}&=&\frac{(E_T^{jet1}e^{\eta^{jet2}}+
                 E_T^{jet2}e^{\eta^{jet2}})}{2(E_p-E_n)}.
\label{eqxg}
\end{eqnarray}

\noindent
 The pseudorapidity is defined by 
   $\eta = -\ln{(\tan\frac{\theta}{2})}$, where
   $\theta$ is the polar angle with respect to the $z$ axis.

%=================================================================

\section{Experimental Procedure}

\subsection{H1 detector}

The data used in this analysis were collected with the 
H1 detector at HERA in the years 1996-97 and correspond to an integrated
luminosity of 19.2 $\rm pb^{-1}$. In these years the HERA collider was operated 
at positron and proton beam energies of $E_e=27.6~\GeV$ and $E_p=820~\GeV$, 
respectively.

A detailed description of the H1 detector can be found elsewhere
\cite{h1detector}. Here only the components 
relevant for the present measurement are briefly described.

The $e^+p$ luminosity  is determined  with a precision of 1.6\% by 
detecting photons from the Bethe-Heitler process 
$e^+p\rightarrow e^+p\gamma$ in the photon detector located at $z=-103$~m.
The electron calorimeter of the luminosity system,
located at $z=-33$~m, is used to detect the positrons
scattered through very small angles (i.e. large $\theta'_e$)
in photoproduction events.
Both detectors are TlCl/TlBr crystal \v{C}erenkov calorimeters with
an energy resolution of $\sigma(E)/E=22\%/\sqrt{E[\GeV]}$.

The Central Tracking Detector (CTD), with  a polar angle coverage of
$25\deg<\theta<155\deg$, is used to measure the trajectories of
charged particles and to reconstruct  the interaction vertex.
The tracking system is surrounded by 
the finely segmented Liquid Argon (LAr) calorimeter~\cite{LAR},
which covers the range in polar angle 
$4\deg<\theta<154\deg$ with full azimuthal acceptance.
The LAr calorimeter consists of an electromagnetic section with lead as absorber, and
a hadronic section with steel as absorber.
The total depth of the LAr calorimeter ranges from 4.5 to 8 
hadronic interaction lengths. Its energy resolution, determined
in test beam  measurements, is $\sigma(E)/E\approx 
12\%/\sqrt{E[\GeV]}\oplus 1\%$ for electrons and   
$\sigma(E)/E\approx 50\%/\sqrt{E[\GeV]}\oplus 2\%$ for charged pions.
The absolute electromagnetic energy scale is known with a precision of 1\%.
The absolute hadronic energy scale for the jet energies used 
in this analysis is known with a precision of 4\%.

The polar angle region $153\deg<\theta<177.8\deg$  is covered by 
a lead/scintillating-fibre calorimeter, the SpaCal \cite{SPACAL},
with  both electromagnetic and hadronic sections. 
For positrons, the energy resolution is
$\sigma(E)/E\approx 7.1\%/\sqrt{E[\GeV]}\oplus 1\%$
and the energy scale uncertainty is less than 2\%.
The positron polar angle is measured with a precision of 1~mrad.
The hadronic energy scale in the SpaCal is known with a precision of 7\%.
A Backward Drift Chamber (BDC) in front of the SpaCal
is used to suppress background from neutral particles which can fake the
scattered positron signal.

The LAr and SpaCal calorimeters are surrounded by a superconducting solenoid
which provides a uniform magnetic field of 1.15~T along the beam direction.

Leading neutrons are detected in the Forward Neutron Calorimeter
(FNC), which consists of interleaved layers of 2~m long lead strips 
and scintillator fibres and is located $107$~m away from
the nominal H1 interaction point in the proton beam direction
(for details see~\cite{H1LN}).
The energy resolution of the calorimeter is $\sigma(E)/E \approx 20\%$
for neutron energies between 300 and 820 $\GeV$.
The absolute hadronic energy scale is known with a precision of 5\%.
Two segmented planes of hodoscopes situated in front of the FNC are used
to veto charged particles. Each plane is constructed of 1~cm thick
hexagonal scintillator tiles, which have the same lateral size as
the calorimeter modules.
The  neutron detection efficiency of the FNC is $(93\pm 5)\%$, the losses
being due to the back-scattering of charged particles from the hadronic shower
caused by the neutron which give signals in the veto hodoscopes.
The acceptance of the FNC is defined by the aperture of the HERA beam line
magnets and is limited to neutron scattering angles of 
$\theta_n \lsim 0.8$~mrad with approximately 30\% azimuthal coverage.

%-----------------------------------------------------------------------------
\subsection{Event selection }

The events used in this analysis are triggered by the coincidence 
of a track in the CTD
with an electromagnetic cluster either in the SpaCal (DIS sample) or
in the electron calorimeter of the luminosity system
(photoproduction sample).
% placed close to the beam line 
A number of selection criteria are applied in order to suppress
     background and to confine the measurements to those regions of
     phase space where the acceptance is large and uniform.

The reconstructed $z$ coordinate  of the event vertex is required to be 
within $\pm 30 \ {\rm cm}$ $(\sim 3 \sigma)$ 
of the mean $z$ position  of the interaction point.

In the photoproduction analysis, the scattered positron energy $E'_e$ 
is limited to the range $0.3<y\approx (1-E'_e/E_e )< 0.65$. 
This cut is defined by the geometrical acceptance of the electron 
calorimeter which also restricts the value of $Q^2$ to be less than
$10^{-2}~\GeV^2$.
To ensure that the effect of radiative corrections in photoproduction is small
and to suppress events in which a Bethe-Heitler event 
and a photoproduction event are superimposed, the energy measured in the  
photon detector of the luminosity system is required to be less than
$1.5~\GeV$.

The final state positron candidates in DIS are required to 
have polar scattering  angles in the range
$156\deg<\theta^{\prime}_e<176\deg$ 
and energies $E^{\prime}_e$ greater than 10~$\GeV$. 
The energy and angle, $E^{\prime}_e$ and $\theta '_e$, are determined from the 
associated SpaCal cluster in
combination with the interaction vertex reconstructed in the Central Tracker.
The  analysis is restricted to the region
$2 < Q^2 < 80~\GeV^2$ and $0.1 < y < 0.7$.
To suppress events with initial state hard photon radiation, as well as
events originating  from  non-$ep$ interactions,  
the quantity $E - p_z$, summed over  
all reconstructed particles including the positron, is required to lie 
between  $35~\GeV$ and $75~\GeV$.
This quantity, which refers to the energy and
longitudinal momentum component of each final
state particle, is expected to be twice the
electron beam energy for contained events.

All events that satisfy the selection cuts are subjected to a jet
search using a cone algorithm
with radius 
% $R=\sqrt{\Delta\eta^2+\Delta\phi^2}<1$~\cite{snowmass}. 
$R=1$~\cite{snowmass}. 
The jet finding is performed using the energies measured in the LAr and 
SpaCal calorimeters
in the $\gamma^*p$ centre-of-mass frame, with transverse 
energies calculated  relative to the $\gamma^*$ axis in that frame.
To ensure that the bulk of the jet energy is contained
within the LAr calorimeter, 
the laboratory pseudorapidity of each jet axis is restricted 
to the region  $-1<\eta_{lab}^{jet}<2$.
Events which have exactly two jets are selected. The transverse energies
of these jets must be above 7~$\GeV$ (first jet) and above 6~$\GeV$ 
(second jet).

For the cross section measurements, events with a leading neutron are 
selected from the inclusive dijet samples 
by requiring a cluster in the FNC with an energy above 500~GeV.
At such high energies the background contribution due to other
neutral particles is negligible.

The final photoproduction data sample contains about 69\,000 dijet events,
of which 
372 events contain a leading neutron with energy $E_{\rm FNC}>500~\GeV$.
In the DIS sample 23\,000 dijet events are selected, of which
213 satisfy the neutron identification criteria with
$E_{\rm FNC}>500~\GeV$.

The kinematic regions within which this measurement is made 
% Event selection criteria 
are summarized in Table 1.

\begin{table}[h]
  \centering
 \label{table10}
  \begin{tabular}{| l | l |}
     \hline 
      &  Kinematic regions   \\ \hline & \\[-0.5cm]
   Photoproduction & $Q^2<10^{-2}~\GeV^2$,~~~$0.3<y<0.65$ \\
   DIS  & $2<Q^2<80~\GeV^2$,~~~$0.1<y<0.7$ \\ \hline 
     & \\[-0.4cm]
   Dijets   & $E_T^{jet1}>7~\GeV$,~~~$E_T^{jet2}>6~\GeV$,~~~$-1<\eta_{lab}^{jet1,2}<2$ \\
   Neutrons & $E_n>500~\GeV$,~~$\theta_n<0.8$ mrad \\  \hline 
  \end{tabular}
 \caption{The kinematic regions within which the cross sections
         are measured.}
\end{table}

%=================================================================

\section{Monte Carlo Models}

%POMPYT~2.6~\cite{POMPYT}, 
%PYTHIA~5.7~\cite{PYTHIA},
%RAPGAP~2.8~\cite{RAPGAP}, 
%DJANGO
%LEPTO
%HERWIG

Monte Carlo samples are used to correct the data for 
inefficiencies, acceptance effects, migrations and
the effects of QED radiation. They are also used to
% account
correct
for hadronization effects in the comparison with NLO QCD  calculations.
Monte Carlo predictions based on several leading order QCD models 
are also compared with the data.

All the Monte Carlo programs generate hard parton-level interactions using the
Born level QCD matrix elements with a
minimum cut-off  on the transverse momentum  of the outgoing partons.
They differ in the assumptions made about the origin of the partons,
which may come from either the incident proton or an exchanged pion,
and in the details of the  hadronization models.
After hadronization, the response of the H1 detector to the events
is simulated in detail and they
are passed through the same analysis chain as is used for the data.

In addition to the models which are frequently used
in studies of inclusive jet production, namely PYTHIA~\cite{PYTHIA} for 
photoproduction and RAPGAP~\cite{RAPGAP} and LEPTO~\cite{LEPTO}  for  DIS, 
models in which the hard interaction proceeds only
via $\pi$-exchange are also used:
the $\pi$-exchange version of RAPGAP for  both photoproduction and DIS
and POMPYT~\cite{POMPYT} for photoproduction.
A model in which a colour neutral system is formed 
non-perturbatively by soft colour interactions (SCI)~\cite{sci}
is also compared with the data. 
This mechanism is implemented in the Monte Carlo program LEPTO.

The PYTHIA event generator  simulates hard 
photon--proton interactions  via resolved and direct photon 
processes. It is used with a minimum value  for the transverse momenta 
of the outgoing partons in the hard interaction process $(\hat{p}_t^{min})$
 of $2~\GeV$.
The GRV-LO parton densities  are used  for the photon~\cite{GRVph} and 
the proton~\cite{GRVpr}. 
The photon flux is calculated in the
Weizs\"acker-Williams  % equivalent photon 
approximation \cite{WWA}.
Higher order QCD radiation effects are simulated using initial and 
final state parton showers in the leading log approximation. 
The subsequent fragmentation follows the Lund string model as implemented 
in JETSET~7.4 \cite{JETSET}. PYTHIA can also simulate  
multi-parton
interactions (MI), which are calculated as LO QCD processes between partons 
from the remnants of the proton and the resolved photon. 
The resulting additional final state partons are 
required to have transverse momenta above a cut-off value of  $1.2~\GeV$. 
It has previously been shown~\cite{inclusivejet} that these additional 
interactions improve considerably the description of inclusive 
jet photoproduction.
This option of PYTHIA is referred to as PYTHIA-MI below.        
The PYTHIA calculation is performed with version 5.7 and cross-checked
with version 6.1.

The program LEPTO~6.5~\cite{LEPTO} generates DIS events.
It is based on leading order electroweak cross sections and takes QCD 
effects into account to order $\alpha_s$. As in PYTHIA, 
higher order QCD effects are simulated using leading log parton 
showers and the final state hadrons are obtained via Lund string 
fragmentation. Higher order electroweak  processes are  
simulated using DJANGO~\cite{DJANGO}, an interface
between LEPTO and HERACLES~\cite{HERACLES}.
% Simulations using LEPTO in this standard mode
%were used to apply radiative corrections to the measured cross sections. 
The LEPTO program allows the simulation of 
soft colour interactions~\cite{sci}, through which the 
production of leading baryons and diffraction-like configurations 
is enhanced via 
non-perturbative colour rearrangements between the outgoing partons.
%the probability of which is 
%proportional to the normalized difference in the generalized areas
%of the string configurations before and after rearrangement.
In the following, the predictions based on this approach are 
denoted LEPTO-SCI.

The program RAPGAP~2.8~\cite{RAPGAP} is a general purpose event generator 
for inclusive and diffractive $ep$ interactions.
In DIS, the RAPGAP simulation includes a contribution from resolved photon
events in which the photon structure
is parameterized according to the SaS-2D~\cite{sas2d} 
parton densities. These give a reasonable description
of inclusive dijet production at low $Q^2$~\cite{lowq2jet}. 
 In the version denoted below as RAPGAP-$\pi$,
the program simulates exclusively the scattering of virtual or real photons
off an exchanged pion. Here, 
the cross section for photon--proton scattering to the final state 
$nX$ takes the form

\begin{equation}
  d\sigma^{\gamma^* p\rightarrow nX}=f_{\pi^+/p}(x_L,t)\cdot 
  d\sigma^{\gamma^*\pi^+\rightarrow X},% \left(\frac{x}{x_L},Q^2\right),
\end{equation}

\noindent
where $f_{\pi^+/p}(x_L,t)$ is the pion flux associated with the beam 
proton and $d\sigma^{\gamma^*\pi^+\rightarrow X}$%(x/x_L,Q^2)$ 
denotes  the cross section 
for the hard photon--pion interaction. 
The  pion flux factor is taken from ~\cite{Holtmann}:

\begin{equation}
 f_{\pi^+/p}(x_L,t)=\frac{1}{2\pi}\frac{g^2_{p\pi n}}{4\pi}
(1-x_L)\frac{-t}{(m_\pi^2-t)^2}
  \exp \left(- R^2_{\pi n} \frac{m^2_\pi-t}{1-x_L}\right),
\end{equation}

\noindent 
where  $m_\pi$ is the pion mass, 
$g^2_{p\pi n}/4\pi=13.6$ is the $p\pi n$ coupling constant,
known from phenomenological analyses of low-energy data~\cite{Tim91}, 
and $R_{\pi n}=0.93~\GeV^{-1}$
% $R_{\pi n}=0.64~GeV^{-1}$
is the radius of the pion-neutron Fock state of the proton~\cite{Holtmann}.
For the  range of $t$ and $x_L$ relevant here, this flux parameterization yields
results very similar to those of the parameterization
of \cite{PovhKop}, which includes the full $t$ dependence expected 
from the pion Regge trajectory
and which is used in our previous analysis \cite{H1LN}.
If not otherwise stated, the GRV-$\pi$-LO~\cite{GRV1} pion structure 
function parameterization is used.

The POMPYT~2.6~\cite{POMPYT} Monte Carlo program is an extension of PYTHIA,
which models colour singlet exchange processes in photoproduction.
For pion exchange processes, POMPYT simulates the scattering of 
real photons off the exchanged pion,
using the pion flux parameterization of eq.(7).
POMPYT yields results very similar to those of RAPGAP-$\pi$.

%%%%%%%%%%%%%%%%%%%%%%%%%%%%%%%%%%%%%%%%%%%%%%%%%%%%%%%%%%%%%%%%%%%%%%%%%

\section{Systematic Uncertainties}

The acceptance of the FNC calorimeter is defined by the interaction 
point and the geometry of the beam guiding magnets and is determined 
using Monte Carlo simulations.
The angular distribution of the neutrons produced in reaction (1)
is sharply peaked in the forward 
direction, and the observed cross section therefore depends critically 
on small inclinations of the incoming proton beam with respect to 
its nominal direction. This effect 
is studied with the help of the Monte Carlo simulations 
described above.  The overall uncertainty 
in the FNC acceptance is estimated to be about 10\%.
The uncertainty in the neutron detection efficiency 
leads to an additional 5\% systematic error, and the
uncertainty of the FNC absolute energy scale to a 6\% 
systematic error. These effects  
contribute to the overall normalization error.

The 4\% uncertainty on the absolute hadronic energy scale 
of the LAr calorimeter 
leads to an uncertainty of about 15\% on the jet cross section.   
This is strongly correlated between data points.

The uncertainty of the acceptance  of the electron calorimeter
of the luminosity system,
in the photoproduction case, amounts to about 6\%. This includes the 
uncertainty on the luminosity measurement of 1.6\% and contributes 
to the overall normalization  error.

The uncertainties on the measurements of the positron energy and angle 
in the SpaCal  lead to 6\% systematic uncertainties in the DIS cross
sections.

As shown below, the models based on the pion-exchange mechanism
describe the data well and are therefore used to estimate 
acceptance and migration corrections.     
These corrections are determined from the POMPYT Monte Carlo simulations
in the photoproduction case and using RAPGAP-$\pi$ for DIS.
The uncertainties of the corrections are estimated from the differences
in the results when other models are used: PYTHIA or RAPGAP-$\pi$
in the photoproduction case and LEPTO or RAPGAP in the DIS case.
The estimated uncertainties are between 10\% and 15\%  for all distributions.

Due to the energy cut in the photon detector, QED radiative
corrections are small~\cite{radcor}  in the photoproduction  case
and are neglected here. 
 For the DIS sample, the QED radiative corrections  amount to less than 
10\%, as evaluated using RAPGAP  interfaced 
to the HERACLES program. The uncertainty arising from the radiative 
corrections is about 5\%.

Finally, a 3\% correlated uncertainty is attributed to the 
trigger efficiencies
as evaluated using other, independent triggers.

The correlated error contributions are only weakly dependent 
on the kinematic variables studied, causing 
a normalization uncertainty of about 20\% on the cross sections in both 
the DIS and  photoproduction cases. The uncorrelated, point-to-point systematic 
uncertainties range from 11\%  to 17\%.

In the figures, the outer error bars represent the quadratic sum of the
point-to-point systematic errors and the statistical errors,
while the inner error bars show the statistical errors.
The normalization error is not shown in the figures, but is included
in the tables as a correlated systematic error.

%%%%%%%%%%%%%%%%%%%%%%%%%%%%%%%%%%%%%%%%%%%%%%%%%%%%%%%%%%%%%%%%%%%%%%%%%%

\section{Results and Discussion}

\subsection{Neutron energy distribution}

\begin{figure}[p]
\epsfig{file=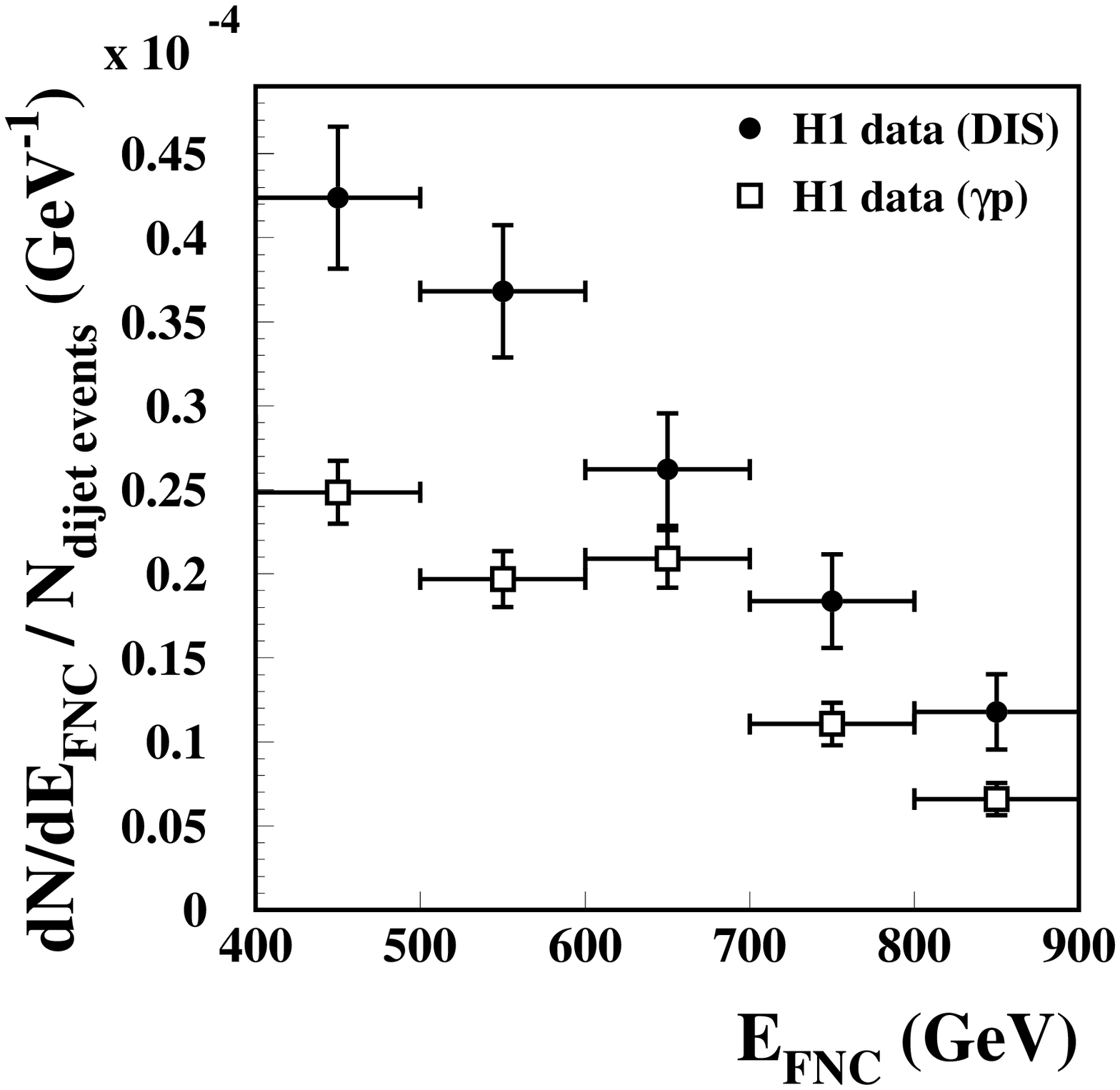,height=82mm}
\hspace*{-8mm}
\epsfig{file=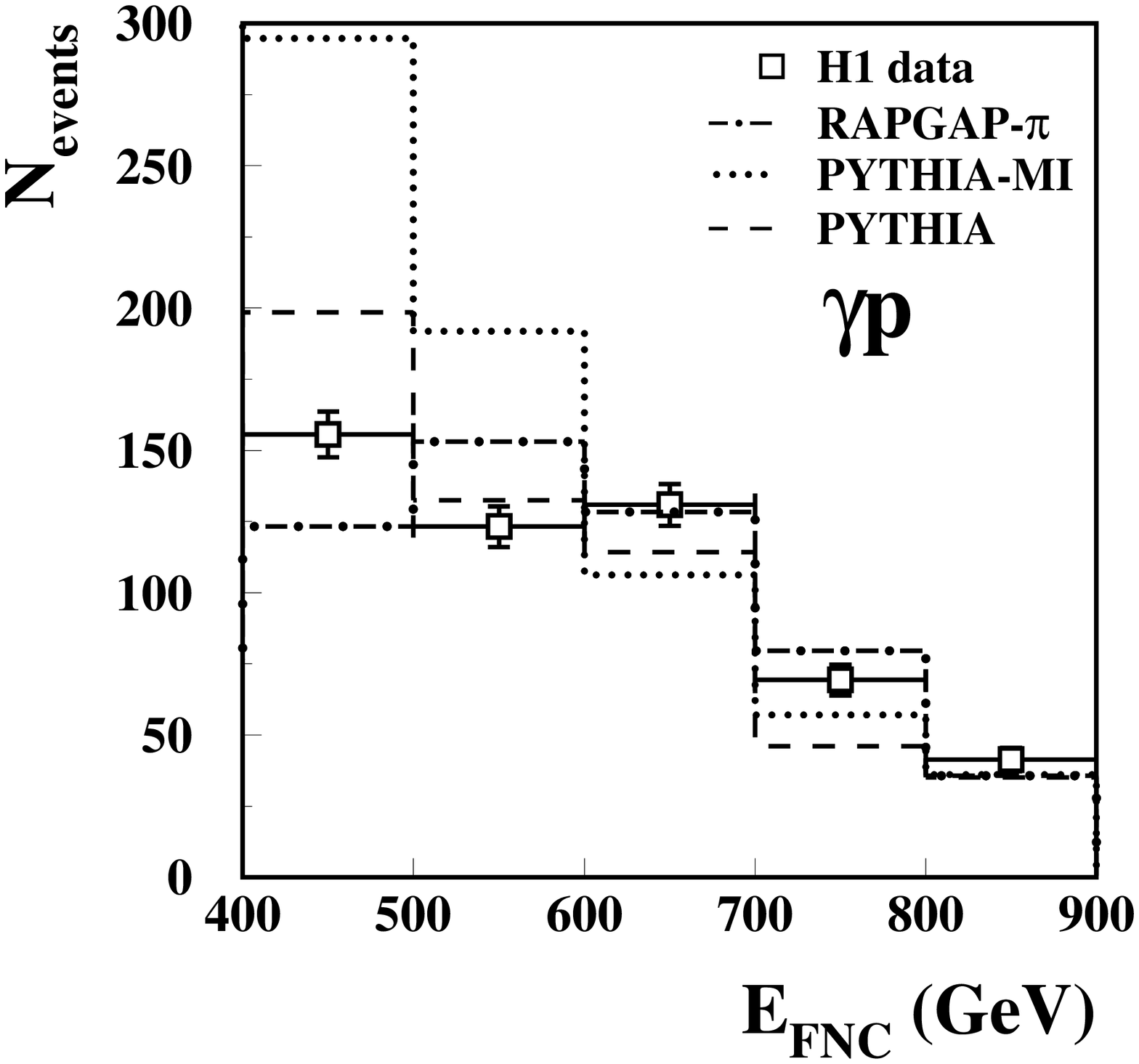,height=82mm}   
\epsfig{file=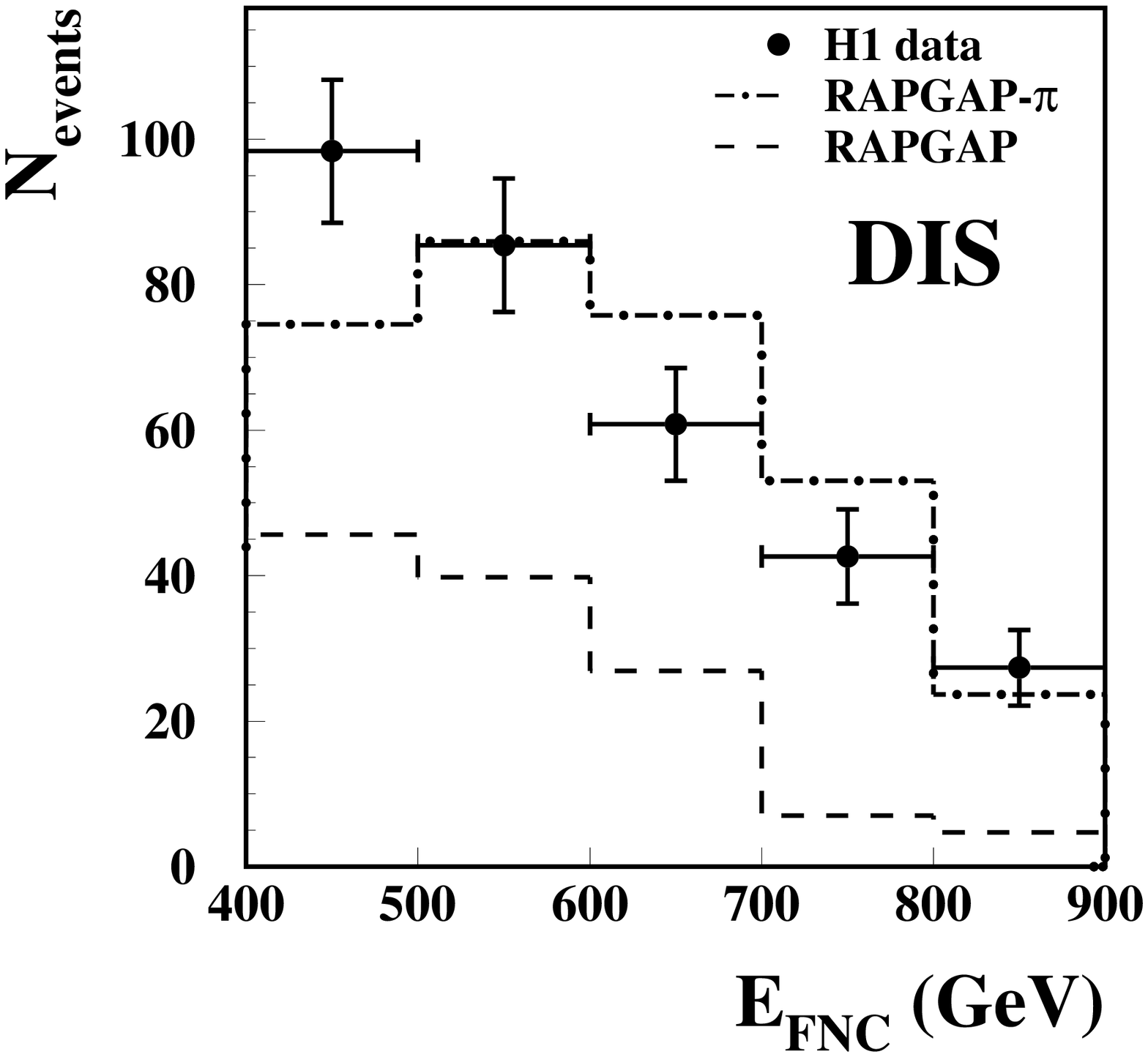,height=82mm}
\hspace*{-8mm}
\epsfig{file=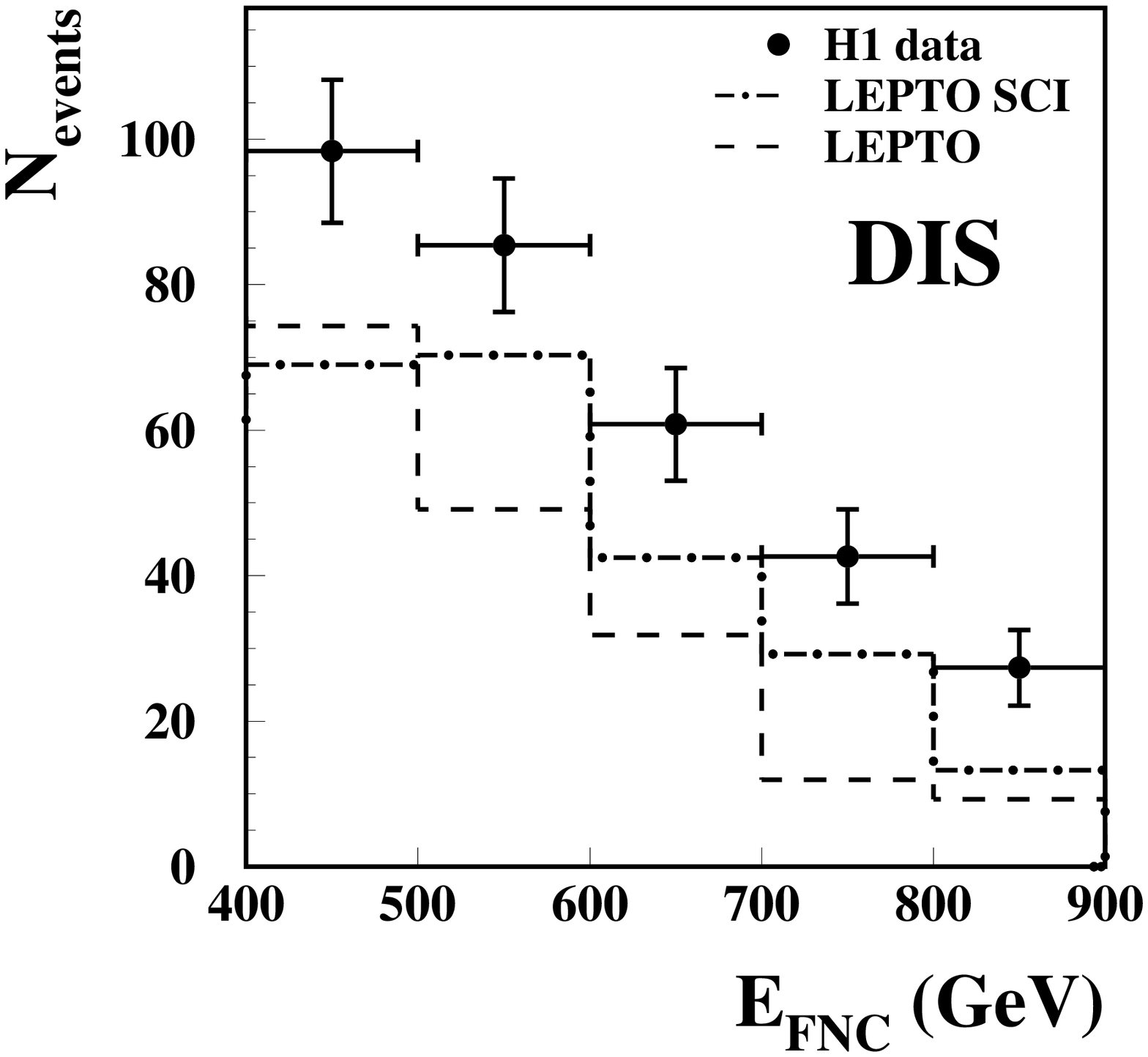,height=82mm}
\caption{Energy distributions observed in the FNC for dijet events: 
(a) Photoproduction data  compared with DIS data, normalized to the 
respective number of dijet events in each inclusive dijet sample;
(b)   The Monte Carlo model predictions for  RAPGAP-$\pi$, PYTHIA 
and PYTHIA-MI,   compared with photoproduction data;
(c,~d)
  The Monte Carlo models RAPGAP (both the $\pi$-exchange and the 
  standard DIS versions) and LEPTO (with and without SCI), 
  compared with DIS data.
  The errors on the data points are statistical only.
  The corresponding kinematic regions for the photoproduction, 
  DIS and dijet selections   are given in Table 1.}
\label{enspectra} 

\vspace*{-148mm} \large\bf \hspace*{25mm} (a) \hspace*{72mm} (b)

\vspace*{77mm}  \large\bf \hspace*{25mm} (c) \hspace*{72mm} (d)

\vspace*{65mm}
\end{figure}

The neutron energy spectrum allows discrimination between different models
for the production of leading neutron events. 
Figure~\ref{enspectra} shows the energy spectra for the 
photoproduction and DIS dijet samples, as measured  in the FNC,
for energies above 400~GeV.
The data are not corrected for efficiencies, acceptance or migration 
between bins.  These effects, however, are taken into account in the 
Monte Carlo simulations.
%%%%%%%%%%%%%%%%%%%%% FIG.2 %%%%%%%%%%%%%%%%%%%%%%%%%%%%%%%%%%%%%%%%%%
%----------------------------------------------------------------
In Fig.~\ref{enspectra}a the photoproduction and DIS data are shown, 
normalized to the total number of events in the corresponding 
inclusive dijet samples. 
% For energies above $400~\GeV$ 
There is a significantly higher fraction of leading neutrons
in DIS dijet events than in photoproduction dijet events.
However, the shape of the energy spectrum is similar in both samples. 

In Figs.~\ref{enspectra}b-d  the photoproduction and DIS data are shown 
together with the predictions from several Monte Carlo models. 
The model predictions are all normalized to the integrated
luminosity of the corresponding data samples.

For simulated neutron energies $E_n>400~\GeV$,
the photoproduction data in Fig.~\ref{enspectra}b are 
reasonably well described in shape 
and magnitude by the $\pi$--exchange model RAPGAP-$\pi$,
as well as by PYTHIA  without multiple 
interactions. If multiple interactions are included, PYTHIA 
fails to describe the data, predicting a rate which is  
too high for neutron energies between 400 and 600~GeV. This observation is in 
 contrast to the inclusive
 jet measurement  (without the requirement of  a leading  neutron), 
 which is   described by PYTHIA only 
 if multiple interactions are included, especially at low jet $E_T$
 \cite{inclusivejet}.

The RAPGAP predictions are compared with DIS data in Fig.~\ref{enspectra}c.
The $\pi$--exchange version describes the shape of the distribution well,
% for $E_n\gsim 500~\GeV$, 
but somewhat overestimates the absolute rate, 
while the rate predicted by the  standard RAPGAP DIS version is too low.  
The LEPTO prediction is also too low, as is shown in 
Fig.~\ref{enspectra}d. The LEPTO predictions are somewhat increased  
for $E_n\gsim 500~\GeV$ if soft colour interactions (LEPTO-SCI) are 
included.

To summarize, the pion exchange models describe  the shape of the 
observed FNC energy spectra well for energies above $500~\GeV$,
as does LEPTO with soft colour interactions.

%----------------------------------------------------------------------
\subsection{Cross section measurements}

In this section, differential cross sections at the hadron level 
are presented for dijet
production in the photoproduction and DIS regimes
for neutron energies $E_n>500~\GeV$, corresponding to the region
in which $\pi$-exchange models give a good description of the 
$E_{\rm FNC}$ distribution.
The data are corrected for detector inefficiencies and migrations 
due to detector resolution  effects using  the Monte Carlo simulations
described in sections 4 and 5.
The results are given in Figures 3--6 and  Tables~2--3.

%%%%%%%%%%%%%%%%%%%%% FIG.3%%%%%%%%%%%%%%%%%%%%%%%%%%%%%%%%%%%%%%%%%%
\begin{figure}[t]
 \centering
\epsfig{file=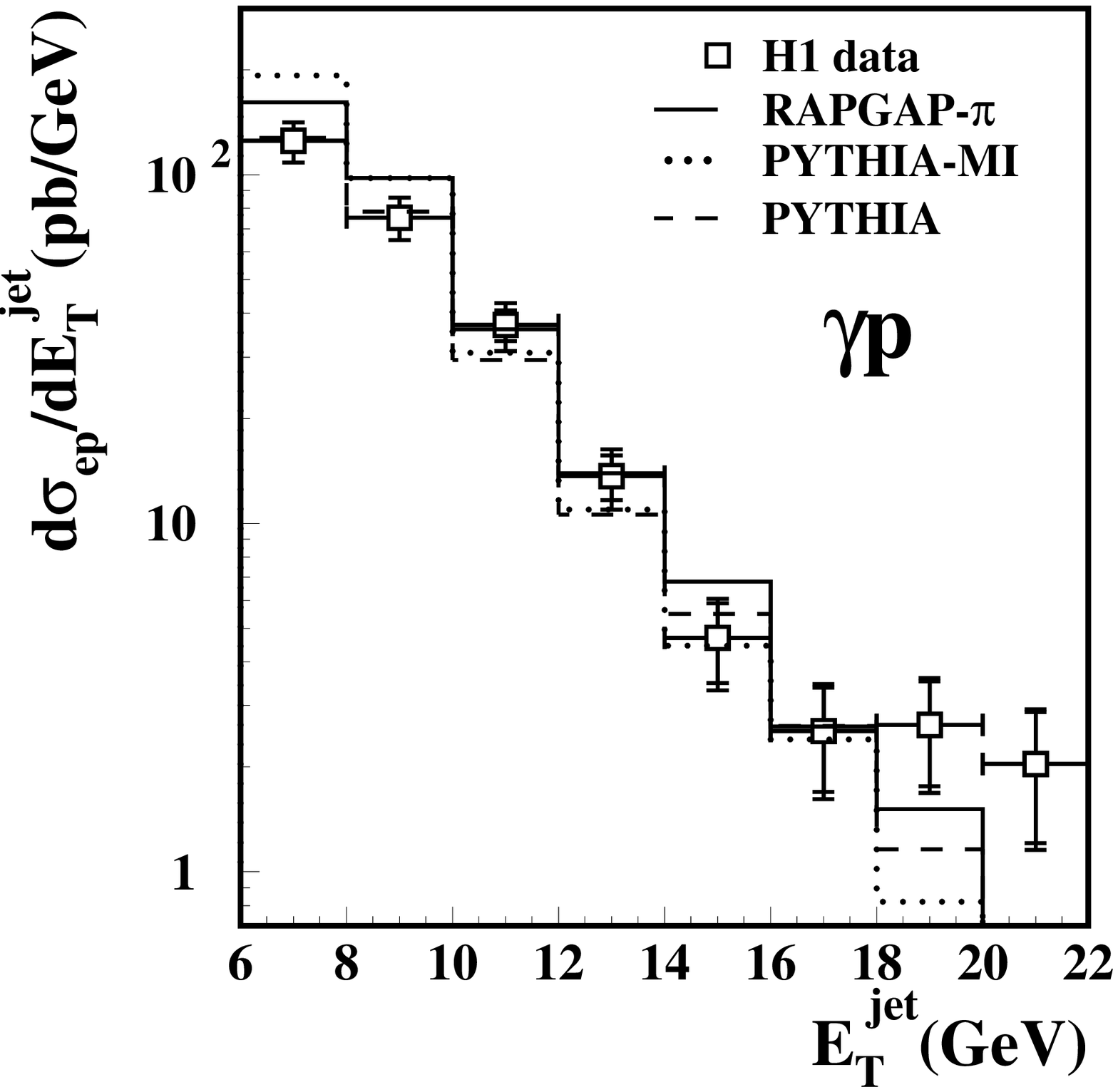,width=79mm}
\epsfig{file=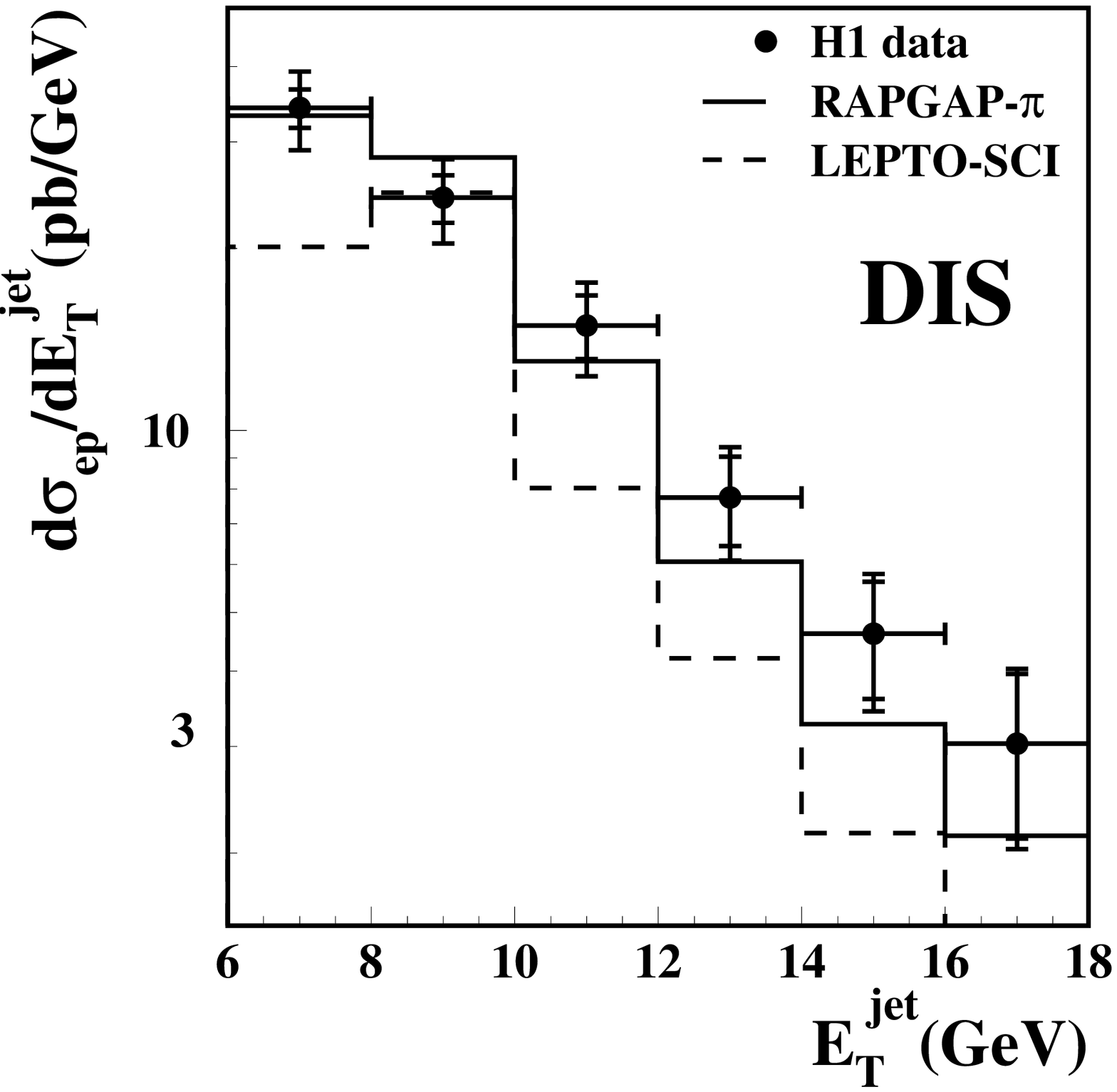,width=79mm}
\epsfig{file=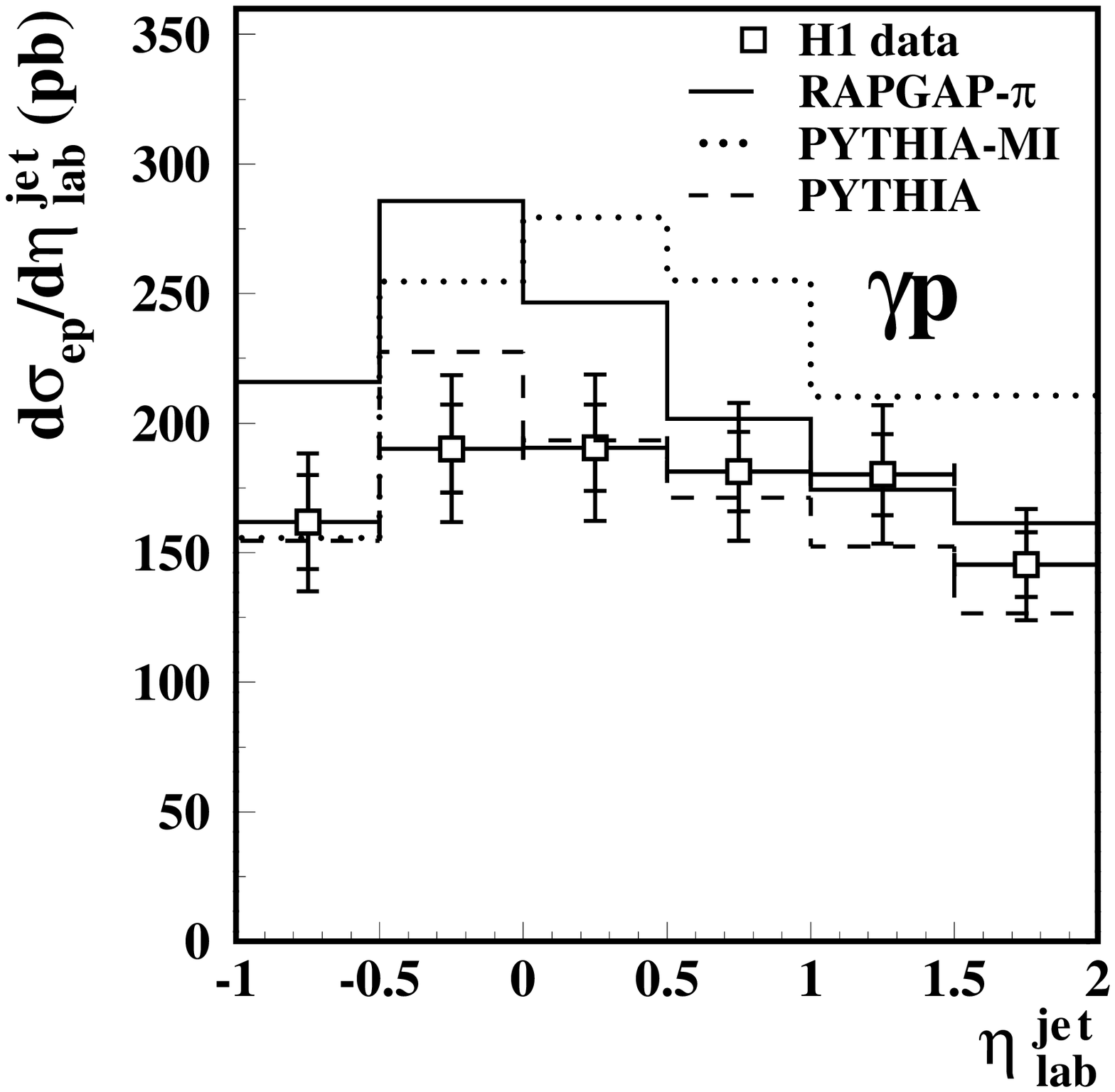,width=79mm}
\epsfig{file=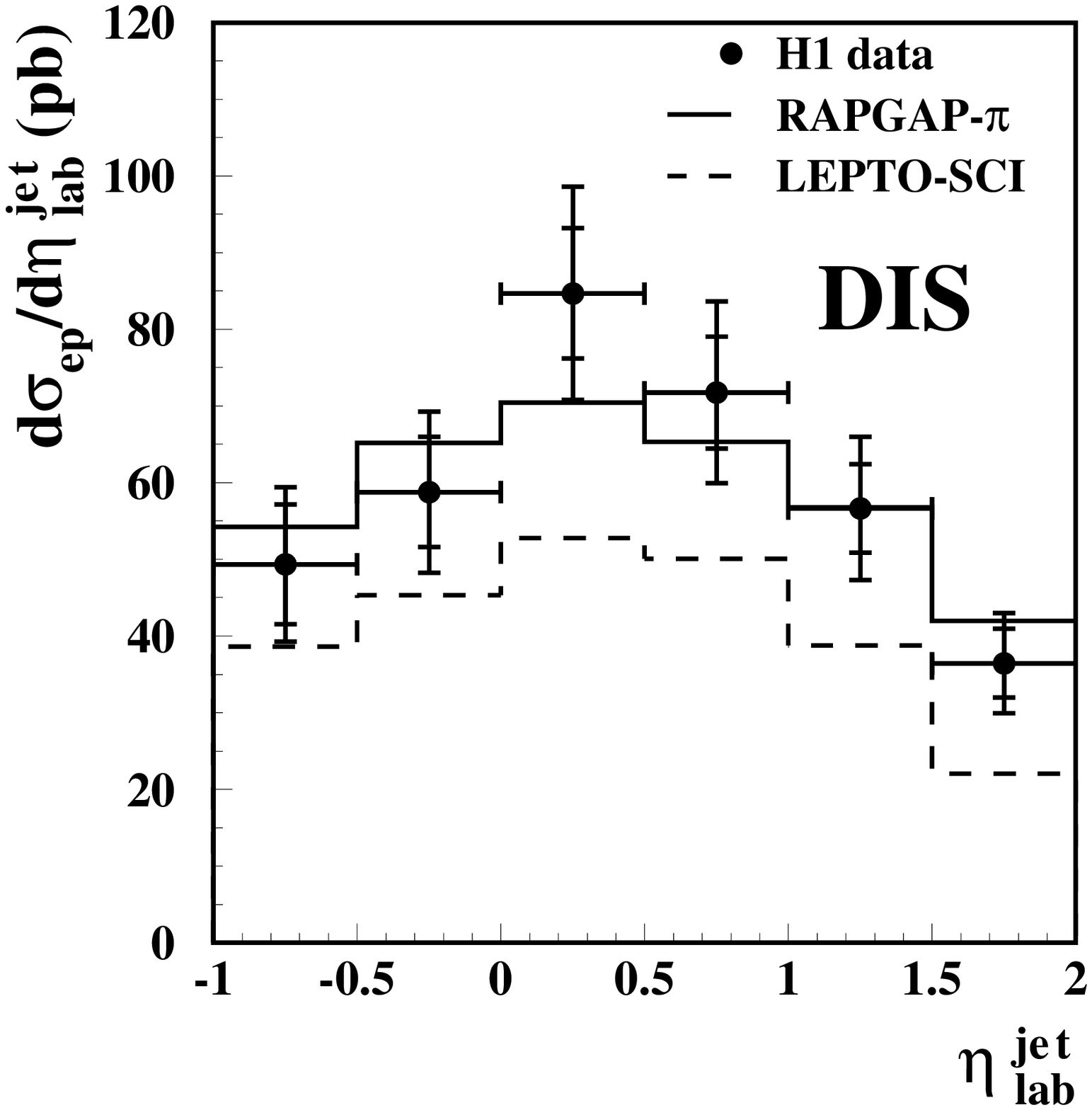,width=79mm}
\caption{The measured differential $ep$ cross sections as a function of
 $E_T^{jet}$  and $\eta_{lab}^{jet}$
for dijet events with  a leading neutron.
The cross section is given for photoproduction (a,c) and 
DIS (b,d) and compared with the Monte Carlo simulations.
Inner error bars show the statistical errors, while
the outer error bars represent the statistical and systematic errors, 
added in quadrature.
The overall normalization uncertainty of 20\% is not shown.
  The kinematic regions within which this measurement is made
  are given in Table 1.}
\label{eteta} 

\vspace*{-142mm} \large\bf \hspace*{-32mm} (a)
\hspace*{72mm} (b) 

\vspace*{77mm}  \large\bf \hspace*{-32mm} (c) \hspace*{72mm} (d)

\vspace*{54mm}

\end{figure}  

%--------------------------------------------------------------------

In Fig.~\ref{eteta} the jet cross sections are shown as a 
function of $E_T^{jet}$ and $\eta_{lab}^{jet}$ for the 
photoproduction and DIS regimes  
using both jets in the event.

Taking the 20\% normalization uncertainty into account,
the data are well described by the $\pi$--exchange model RAPGAP-$\pi$
in both  DIS and  photoproduction.
However,  PYTHIA without multiple interactions,
which does not include pion exchange,
also provides a good description of the photoproduction data. 
The inclusion of multiple interactions in PYTHIA
results in a predicted cross section which  is too high
 for low values of $E_T^{jet}$ and for 
 values of $\eta_{lab}^{jet}>-0.5$. 
It is also seen from Figs.~\ref{eteta}b and \ref{eteta}d  
that standard DIS processes, as simulated by 
the LEPTO program, tend to lie below the data,
even if soft colour interactions are included.

\begin{figure}[t]
 \centering
\epsfig{file=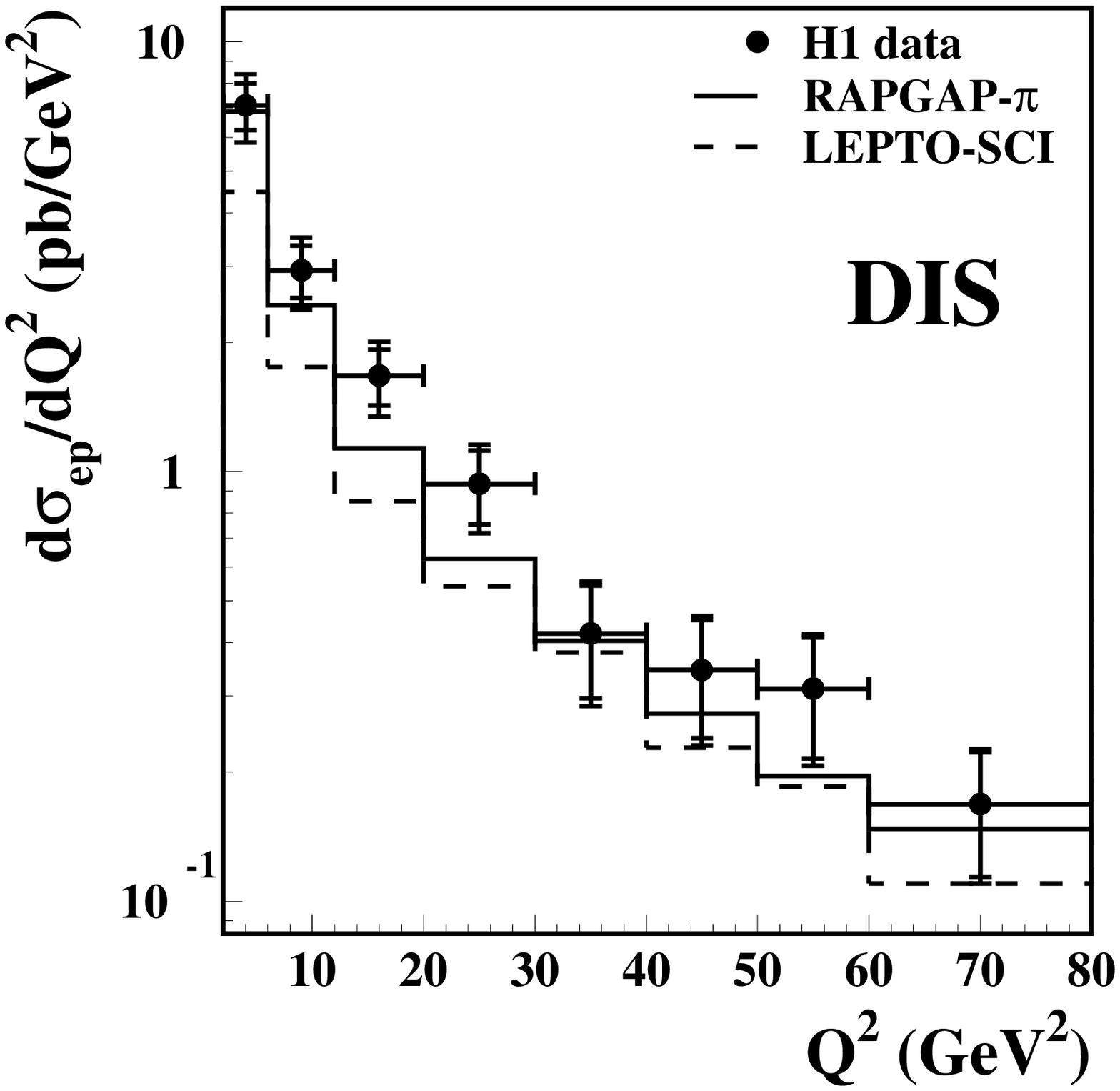,width=88mm}
\caption{The measured deep inelastic $ep$ cross section as a function of
$Q^2$ for dijet events with a leading neutron.
Inner error bars show the statistical errors, while
the outer error bars represent the statistical and systematic errors, 
added in quadrature.
The overall normalization uncertainty of 20\% is not shown.
Predictions from Monte Carlo simulations are compared with the measurements.
 The kinematic regions within which this measurement is made
 are given in Table 1.}
\label{q2}
\end{figure}

The measured $Q^2$ dependence of the dijet  DIS cross section is shown in 
Fig.~\ref{q2}, together with predictions from Monte Carlo
simulations. The $\pi$-exchange
version of RAPGAP describes the measured distribution fairly well,
whereas LEPTO with SCI reproduces the  shape of the distribution but
 yields a lower prediction over the whole $Q^2$ range.

%%%%%%%%%%%%%%%%%%%%% FIG.4 %%%%%%%%%%%%%%%%%%%%%%%%%%%%%%%%%%%%%%%%%%

%----------------------------------------------------------------------

\begin{figure}[t]
 \centering
   
\epsfig{file=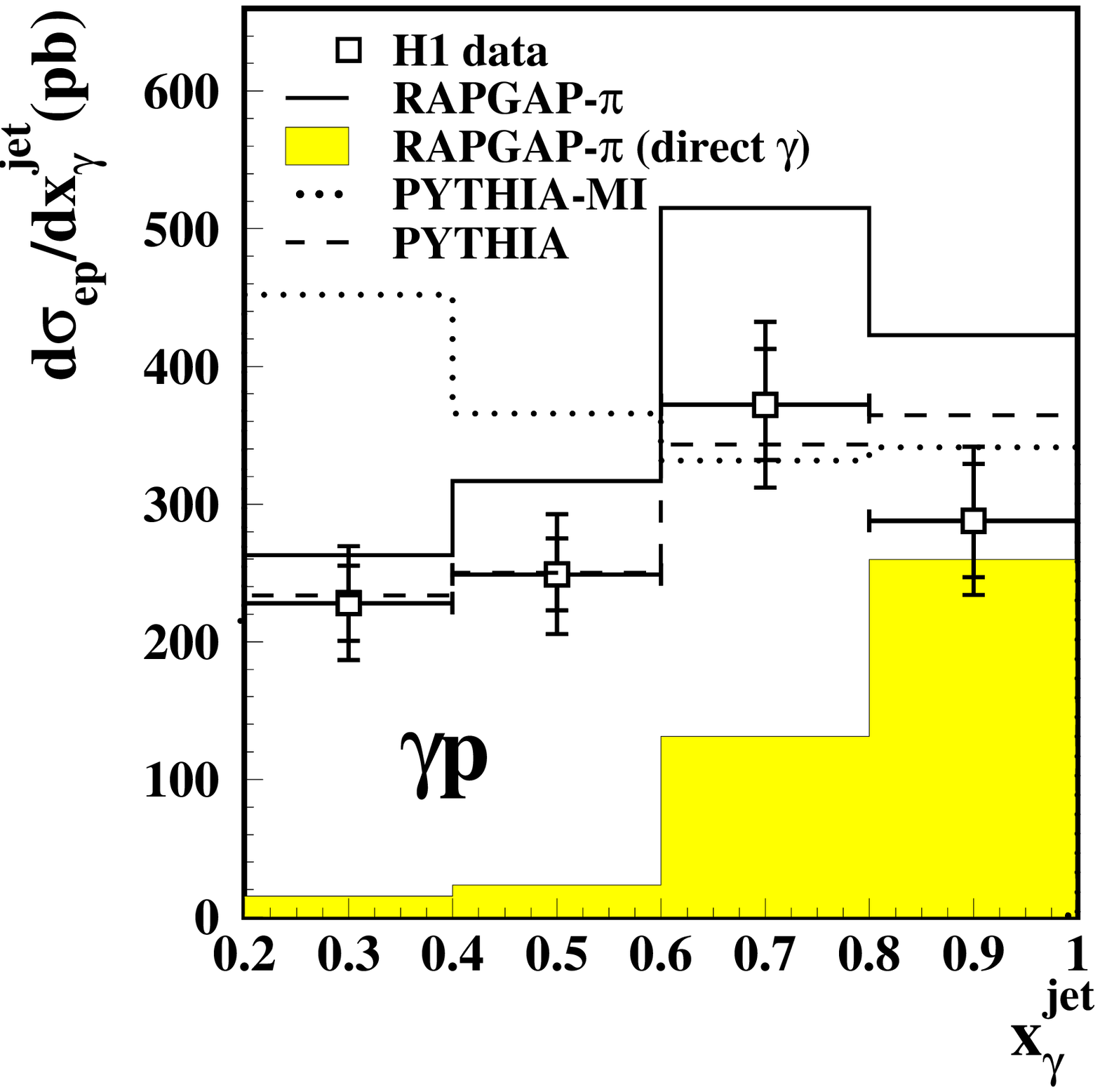,width=79mm}
\epsfig{file=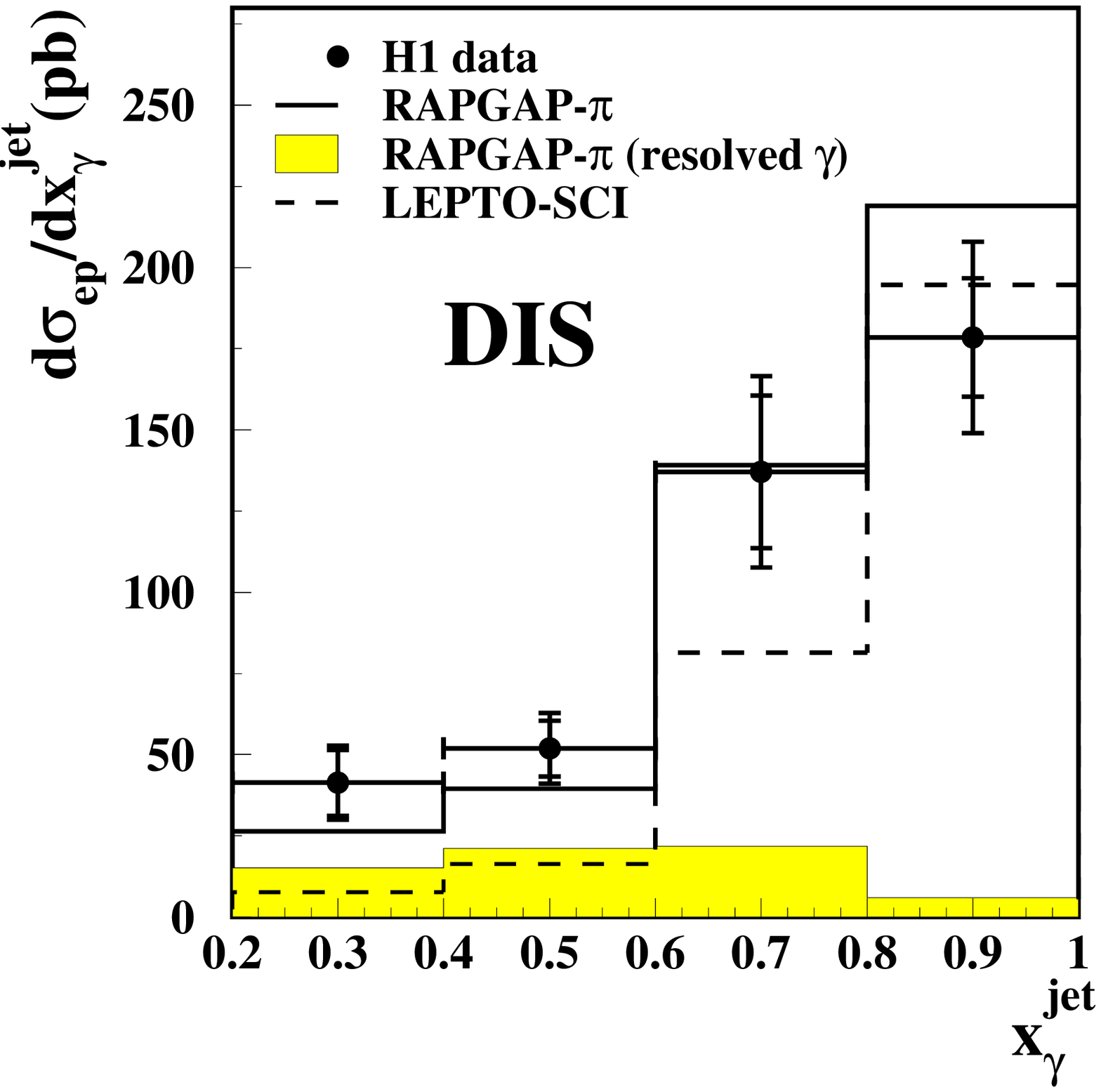,width=79mm}
\caption{The measured differential $ep$ cross sections as a function of
 $x_\gamma^{jet}$ for dijet events with  a leading neutron.
The cross section is given for photoproduction (a) and
DIS (b) and compared with the Monte Carlo simulations.
Inner error bars show the statistical errors, while
the outer error bars represent the statistical and systematic errors,
added in quadrature.
The overall normalization uncertainty of 20\% is not shown.
  The kinematic regions within which this measurement is made
  are given in Table 1.}
 \label{xgamma}

\vspace*{-106mm}  \large\bf \hspace*{41mm} (a) \hspace*{70mm} (b)

\vspace*{103mm}

\end{figure}

\begin{figure}[h]
 \centering
\epsfig{file=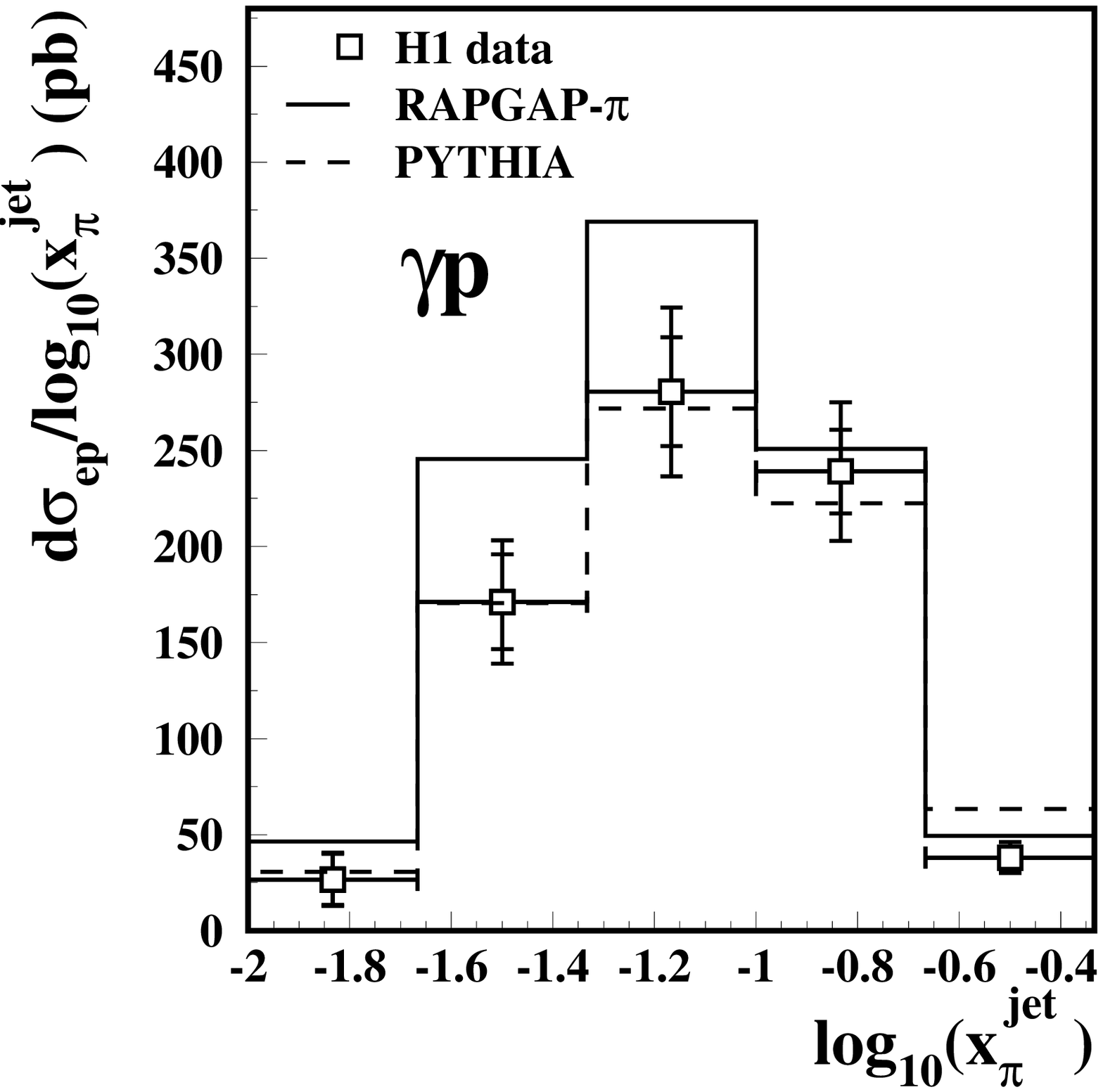,width=79mm}
\epsfig{file=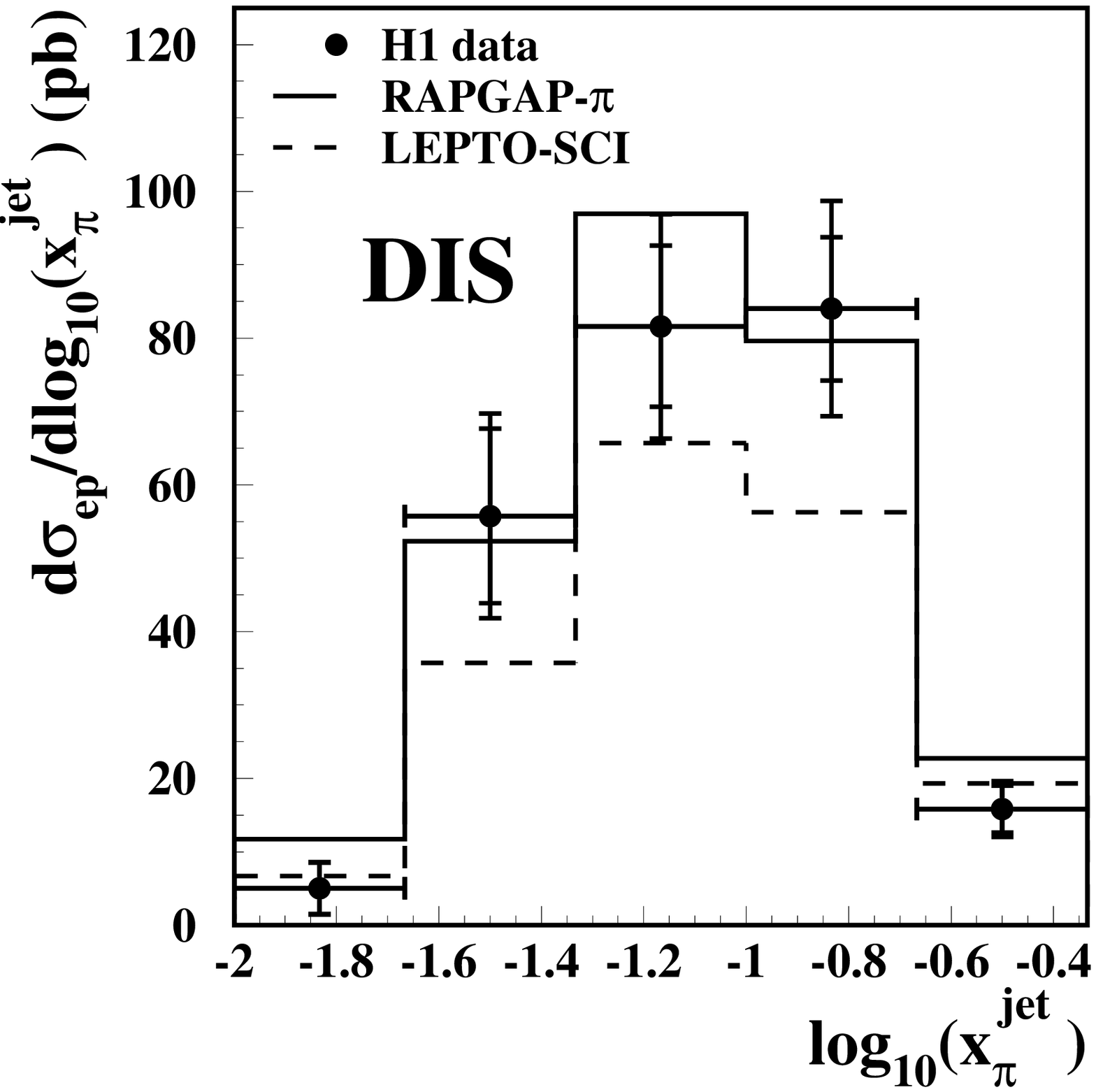,width=79mm}
\caption{The measured differential $ep$ cross sections as a function of
 $x_\pi^{jet}$ for dijet events with  a leading neutron.
The cross section is given for photoproduction (a) and 
DIS (b) and compared with the Monte Carlo simulations.
Inner error bars show the statistical errors, while
the outer error bars represent the statistical and systematic errors,
added in quadrature.
The overall normalization uncertainty of 20\% is not shown.
  The kinematic regions within which this measurement is made
  are given in Table 1.}
\label{xpi}

\vspace*{-106mm}  \large\bf \hspace*{42mm} (a) \hspace*{70mm} (b)

\vspace*{100mm}

\end{figure}

The dependence of the dijet cross section on the fractional 
momenta $x_\gamma^{jet}$ and $x_\pi^{jet}$, 
determined according to eq.(5), is shown in Figs.~\ref{xgamma} and 
\ref{xpi}. 
The measured $x_\gamma^{jet}$ distribution in the photoproduction regime,
together with the RAPGAP-$\pi$ and PYTHIA  model predictions shown in
Fig.~\ref{xgamma}a,
clearly demonstrate the large contribution of resolved photon processes.
The shape of the distribution is well described by PYTHIA and  RAPGAP-$\pi$.
PYTHIA with multiple interactions  predicts too high a cross section 
at $x_\gamma^{jet}<0.6$.
In the DIS regime, as is clear from the $x_\gamma^{jet}$ distribution
shown in Fig.~\ref{xgamma}b, 
direct photon interactions dominate. However, a small fraction  
($\sim$ 15\%) of  resolved photon interactions is
necessary to fully describe the data with the RAPGAP-$\pi$ simulation.
The LEPTO-SCI model, which does not include resolved photon processes,
provides a poor description of the shape of the distributions.

The $x_\pi^{jet}$ distributions in the photoproduction and the DIS samples,
shown in Figs.~\ref{xpi}a and ~\ref{xpi}b, respectively, 
are similar in shape. The $\pi$--exchange model
RAPGAP-$\pi$, and the PYTHIA and LEPTO-SCI models, provide a fair 
description of the data.

%%%%%%%%%%%%%%%%%%%%% FIG.5 %%%%%%%%%%%%%%%%%%%%%%%%%%%%%%%%%%%%%%%%%%

%%%%%%%%%%%%%%%%%%%%% FIG.6 %%%%%%%%%%%%%%%%%%%%%%%%%%%%%%%%%%%%%%%%%%

For the RAPGAP-$\pi$ predictions shown in Figs.~3 to 6 the GRV-$\pi$-LO
parameterization of the pion parton distribution functions (PDFs)
is used. However, within the 20\% normalization uncertainty, 
a similar quality of description is provided if other parameterizations
of the pion PDFs~\cite{other1,other2} are used.

%%%%%%%%%%%%%%%%%%%%%%  FIGURE 7  %%%%%%%%%%%%%%%%%%%%%%%%%%

\begin{figure}[p]
 \centering
\epsfig{file=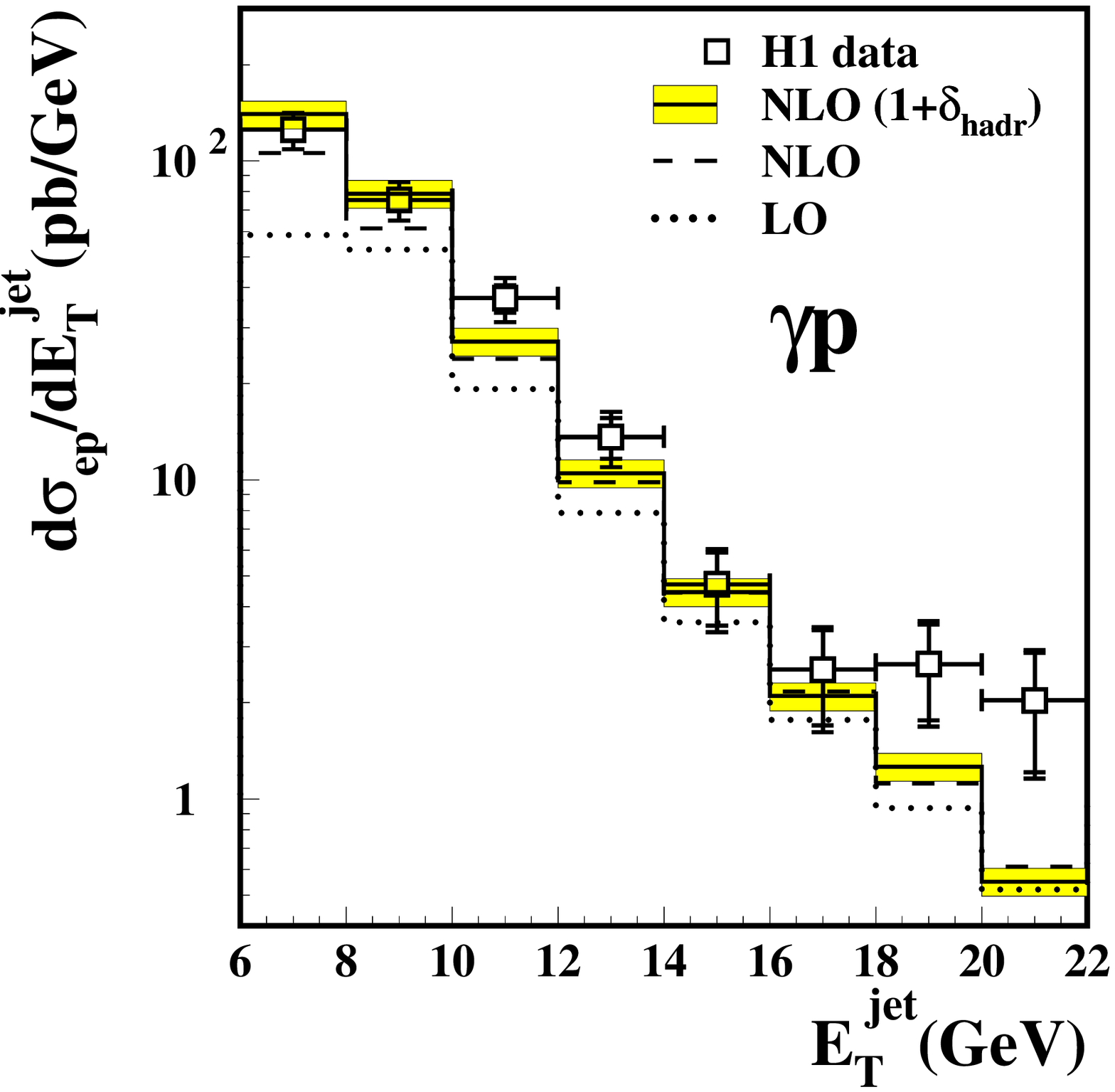,height=82mm}
\epsfig{file=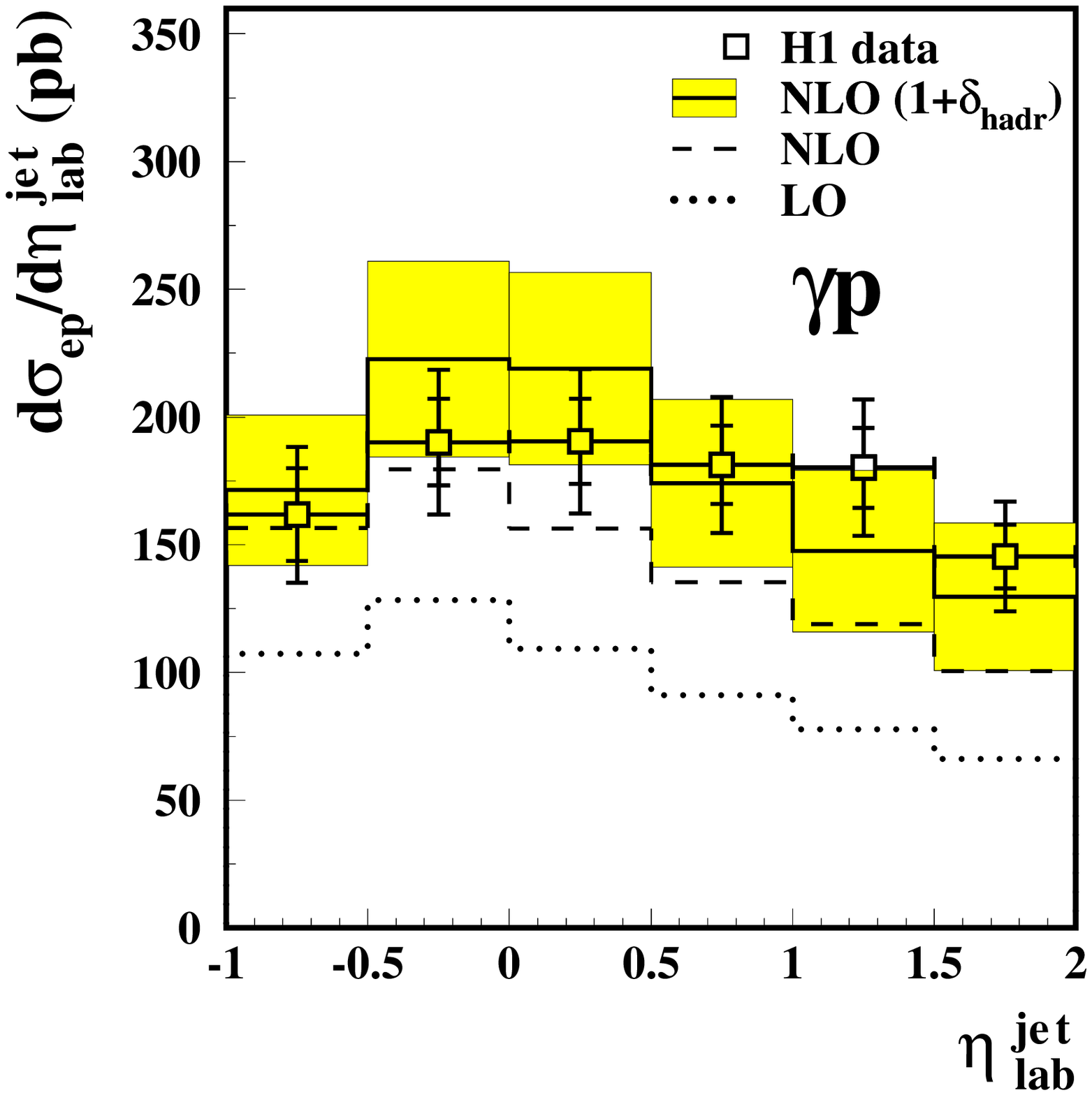,height=82mm}
\epsfig{file=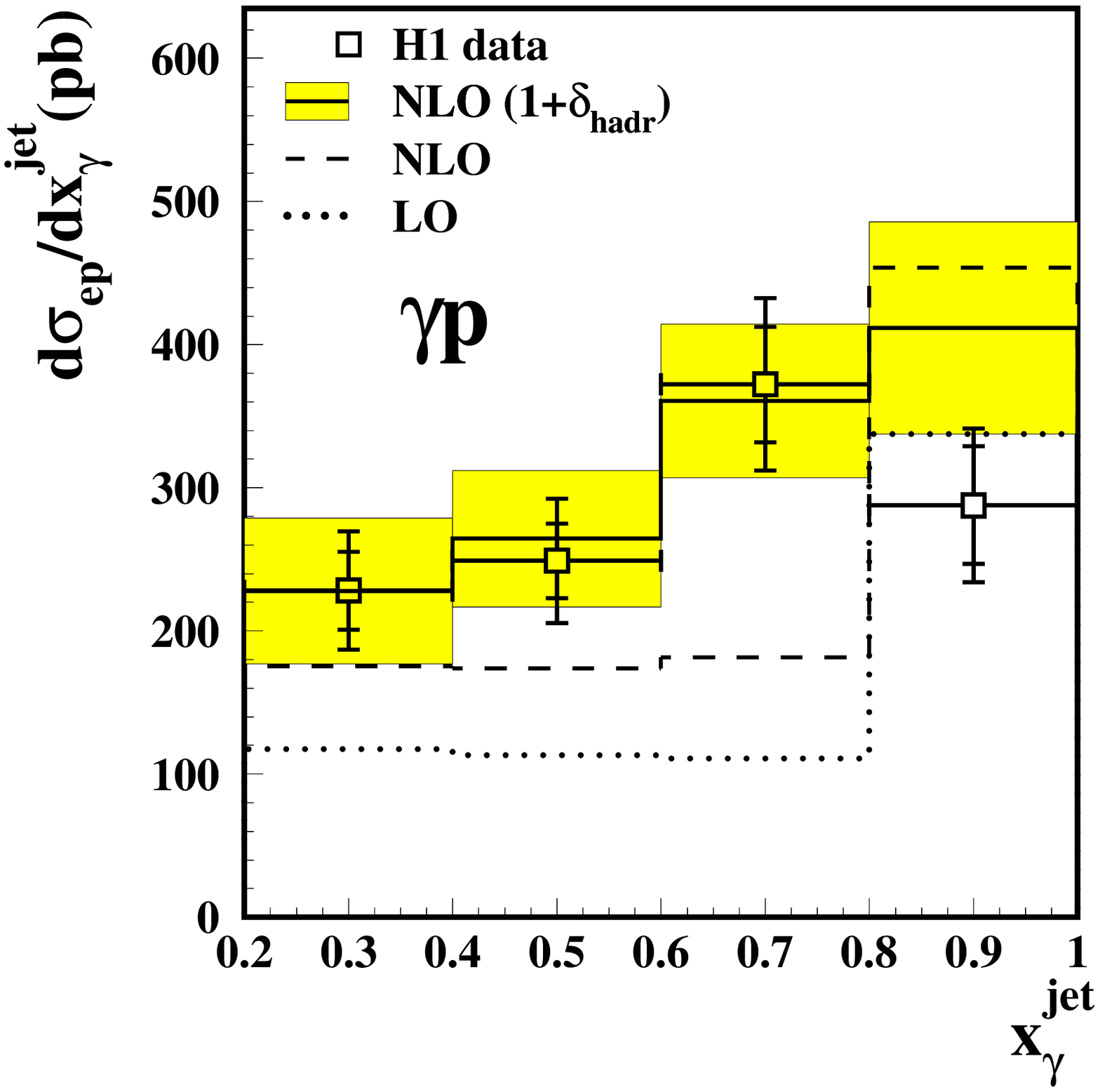,height=82mm}
\epsfig{file=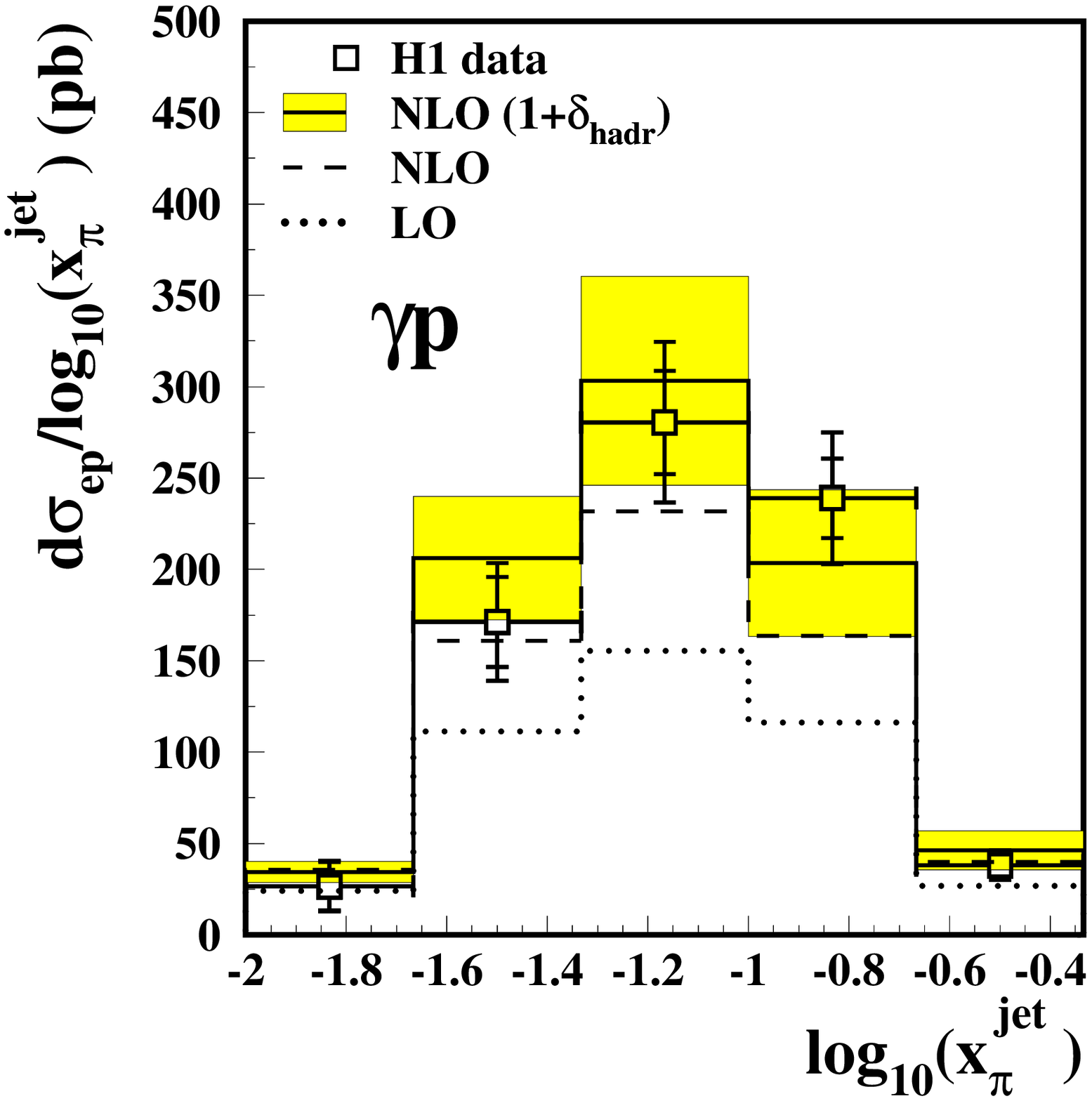,height=82mm}
\caption{The measured differential $ep$ photoproduction cross section
as a function of
$E_T^{jet}$, $\eta_{lab}^{jet}$, $x_\gamma^{jet}$ and $x_\pi^{jet}$
for dijet events with a leading neutron.
Inner error bars show the statistical errors, while
the outer error bars represent the statistical and systematic errors,
added in quadrature.
The overall normalization uncertainty of 20\% is not shown.
LO, and NLO QCD predictions~\cite{Klasen}
 before and after correction for hadronization effects, 
are compared with the measurements. The shaded bands show
the quadratic sum of 
the factorization and renormalization scale uncertainties of 
the NLO predictions and of the uncertainty
due to the hadronization corrections, $1+\delta_{hadr}$.
  The kinematic regions within which this measurement is made
  are given in Table 1.}
\label{nlocalvh} 
\end{figure}  

%*************************************************************************
\subsection{Comparison of  the photoproduction cross section with NLO QCD}

The parton level cross sections for dijet photoproduction 
in photon--pion and photon--proton collisions are calculated 
in QCD at both leading and next-to-leading order \cite{Klasen}.
In the latter case, 
%
%The  LO predictions are obtained using LO matrix elements with the one-loop
%formula for $\alpha_s$. % and the same value of $\Lambda_{QCD}$ of ????.
%The NLO analytical calculations include the tree-level Born matrix elements,
%the virtual corrections with one internal loop, and the real corrections 
%resulting from the
%radiation of a third particle in the initial or final state.
infrared and collinear singularities are cancelled using the phase space
slicing method with an invariant mass cut-off.
The renormalization and factorization scales are defined to be the maximum
transverse energy of the outgoing partons.
The scale uncertainty  amounts to approximately 15\%
on average, but is significantly larger  (up to 30\%)
for low  $x_\gamma^{jet}$ and high $\eta_{lab}^{jet}$, as estimated by varying
the scales by factors of 0.5 and 2.
The photon flux is calculated using the
Weizs\"acker-Williams
% equivalent photon 
approximation~\cite{WWA}.
The light-cone form factor~\cite{Holtmann} is used in the pion flux 
with the same parameters
as for the RAPGAP-$\pi$ Monte Carlo predictions.
In the calculations, the GRV parameterizations  for the parton distribution 
functions  are used for both the photon and the pion.
A cone algorithm with radius $R=1$ is used in the definition of jets.

Since the QCD calculations refer to jets of partons, whereas the 
measurements refer to jets of hadrons, the predicted cross sections 
are corrected to the hadron level using factors evaluated from
the LO Monte Carlo programs described in section~4. 
 The hadronization correction factor, $(1+\delta_{hadr})$, is
defined as the ratio of the cross section obtained with jets
reconstructed from hadrons  
to that using jets reconstructed at the parton level after the 
generation of parton showers.
The corrections are calculated by taking an average of the results from 
two different Monte Carlo models (POMPYT and RAPGAP-$\pi$). 
The uncertainty of these corrections
is taken to be half the difference between the results obtained from the
two models, which is typically smaller than 5\%.
The hadronization corrections have a tendency to increase the 
calculated NLO cross section 
at low $E_T^{jet}$ (by approximately 30\% for $E_T^{jet}=7~\GeV$) 
and to decrease the cross section at 
high $E_T^{jet}$ (by approximately $\rm -$10\%  for $E_T^{jet}=20~\GeV$).
As a function of $x_\gamma^{jet}$, the hadronization 
corrections increase the cross section by about 
25\% at the lowest $x_\gamma^{jet}$, by about 100\% for
$x_\gamma^{jet}=0.8$, and are close to zero  for $x_\gamma^{jet}=1.$
The hadronization corrections show only a weak dependence
on $\eta_{lab}^{jet}$ and $x_\pi^{jet}$.

The LO and NLO calculations are shown with the  
measured cross sections in Fig.~\ref{nlocalvh}. 
After corrections for hadronization, there is good agreement between 
the NLO calculations and the measurements.
The LO and NLO predictions without hadronization corrections do not 
describe the data.

%%%%%%%%%%%%%%%%%%%%%%%%%%%%%%%%%%%%%%%%%%%%%%%%%%%%%%%%%%%%%%%%%%%%%%
\subsection{Ratios of leading neutron to inclusive dijet cross sections}

%%%%%%%%%%%%%%%%%%%%%%%%%%%%%%%%%%%%%%%%  Figure 8  %%%%%%%%%%%%%%
\begin{figure}[p]
\epsfig{file=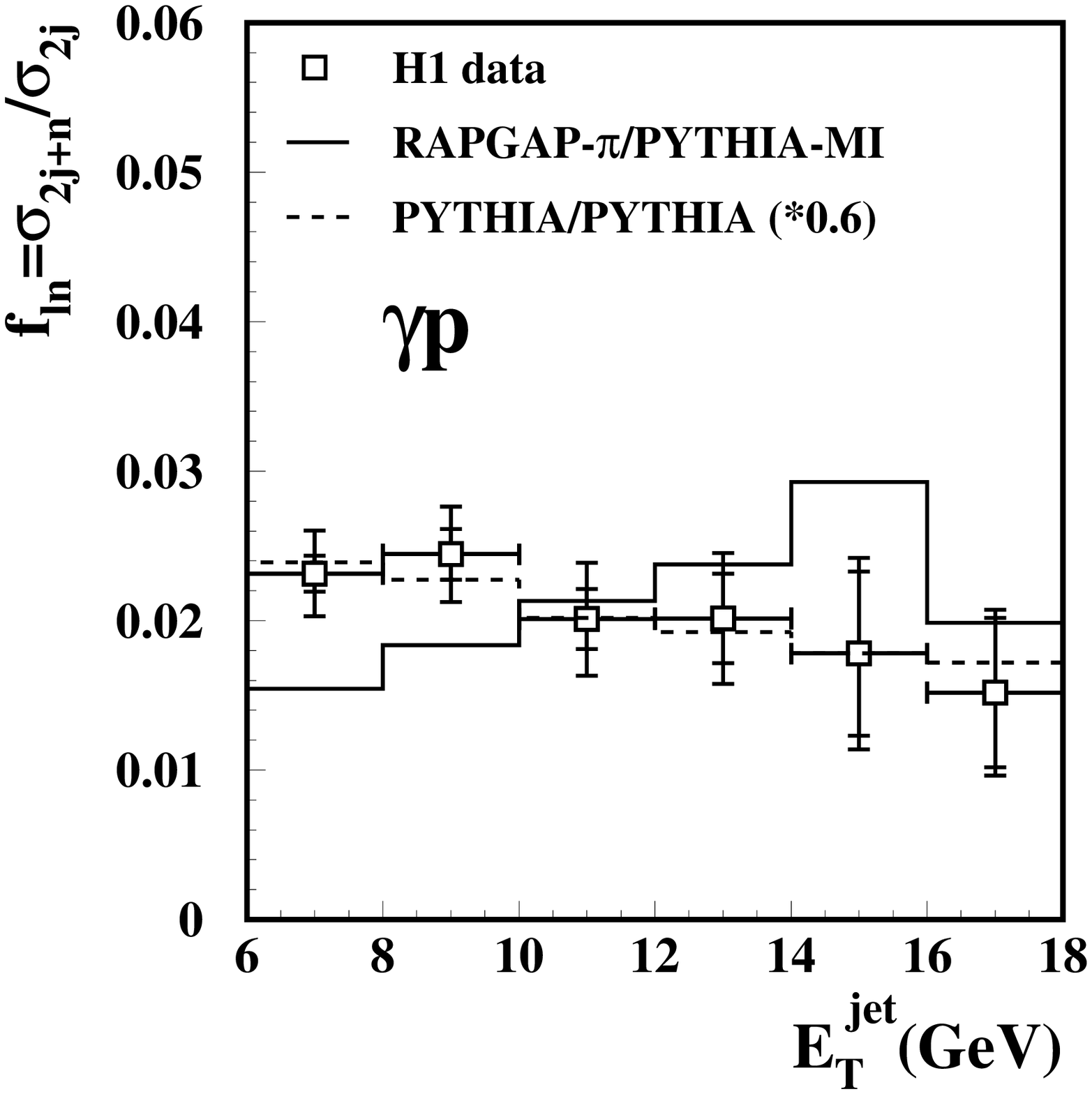,width=79mm}
\epsfig{file=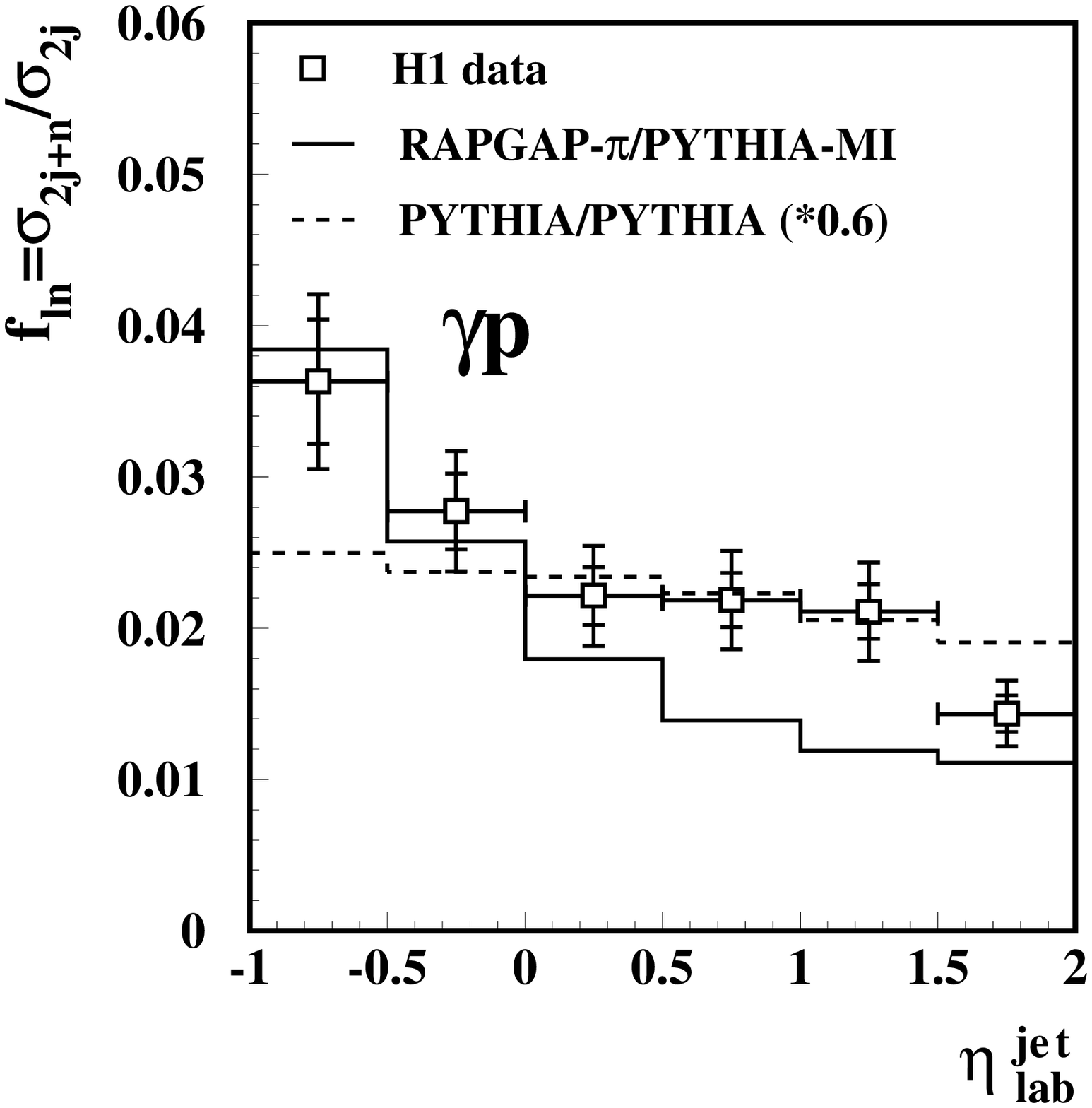,width=79mm}
\epsfig{file=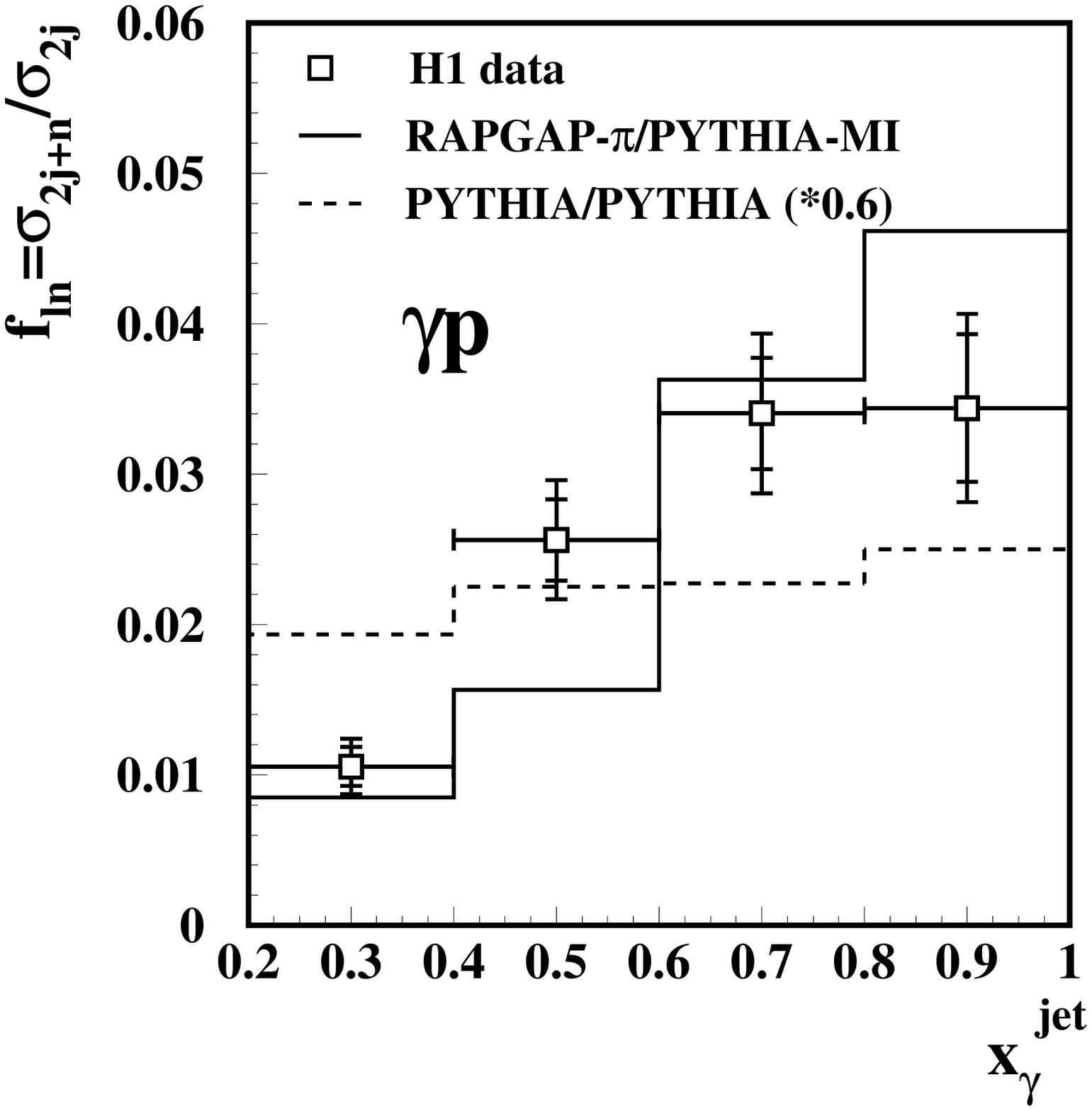,width=79mm}
\epsfig{file=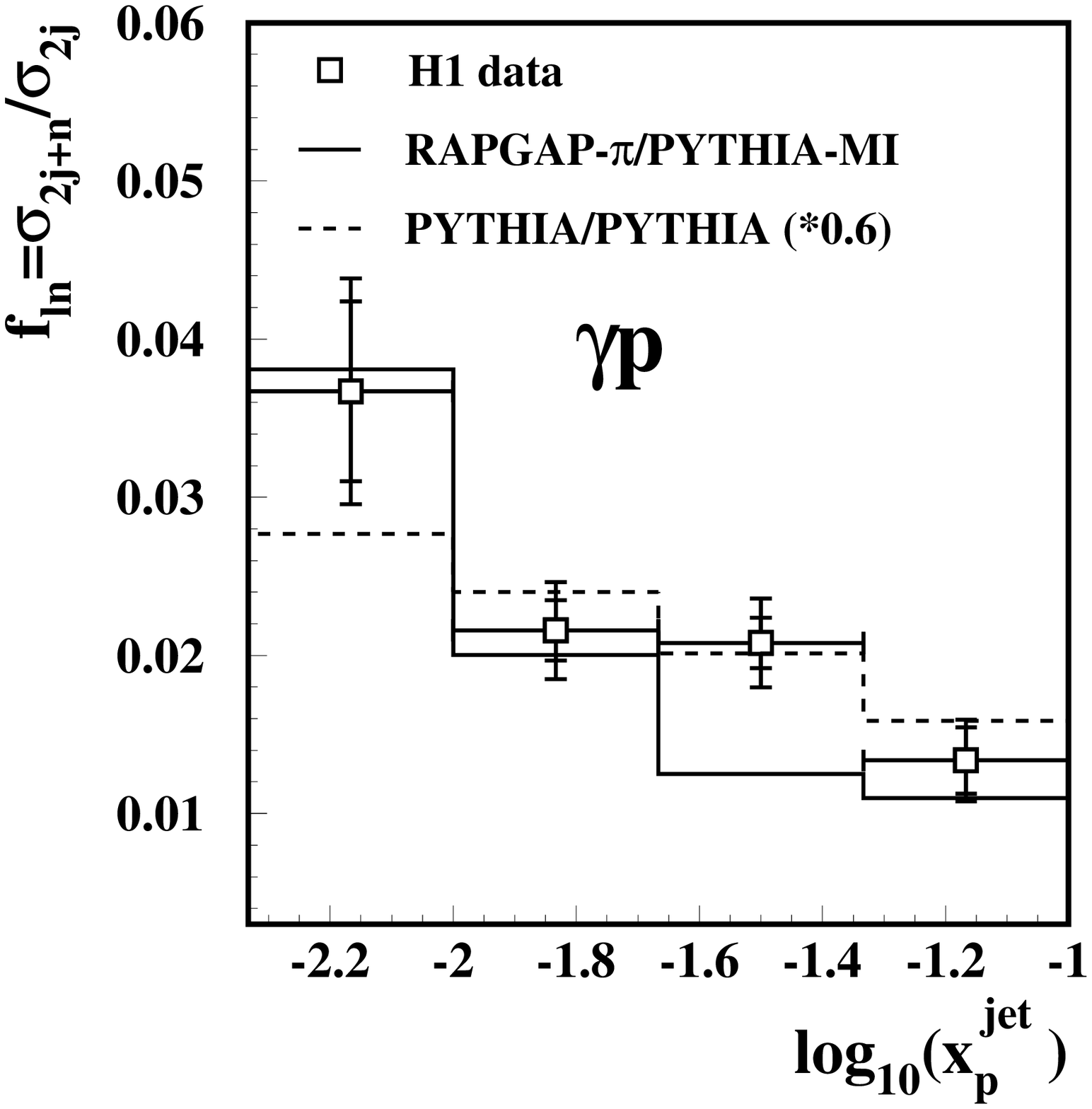,width=79mm}
\caption{The ratio of the cross section for dijet photoproduction
with a leading neutron to that for inclusive dijet photoproduction,
as a function of 
$E_T^{jet}$, $\eta_{lab}^{jet}$, $x_{\gamma}^{jet}$  and $x_p^{jet}$.
Inner error bars show the statistical errors, while
the outer error bars represent the statistical and systematic errors,
added in quadrature.
The overall normalization uncertainty of 13\% is not shown.
Monte Carlo predictions for the ratios are obtained by using
either RAPGAP-$\pi$ for the leading neutron cross sections
and PYTHIA-MI for the inclusive cross sections, or by
using PYTHIA in both cases.
The PYTHIA prediction without multiple interactions has been 
scaled by a factor 0.6 to ease the shape comparison.
  The kinematic regions within which this measurement is made
  are given in Table 1.}
\label{ratio1} 

\vspace*{-157mm} \large\bf \hspace*{21mm} (a)
\hspace*{72mm} (b) 

\vspace*{77mm}  \large\bf \hspace*{21mm} (c) 
\hspace*{72mm} (d)

\vspace*{81mm}

\end{figure}  

%%%%%%%%%%%%%%%%%%%%%%% Figure 9 %%%%%%%%%%%%%%%

\begin{figure}[h]
 \centering
\epsfig{file=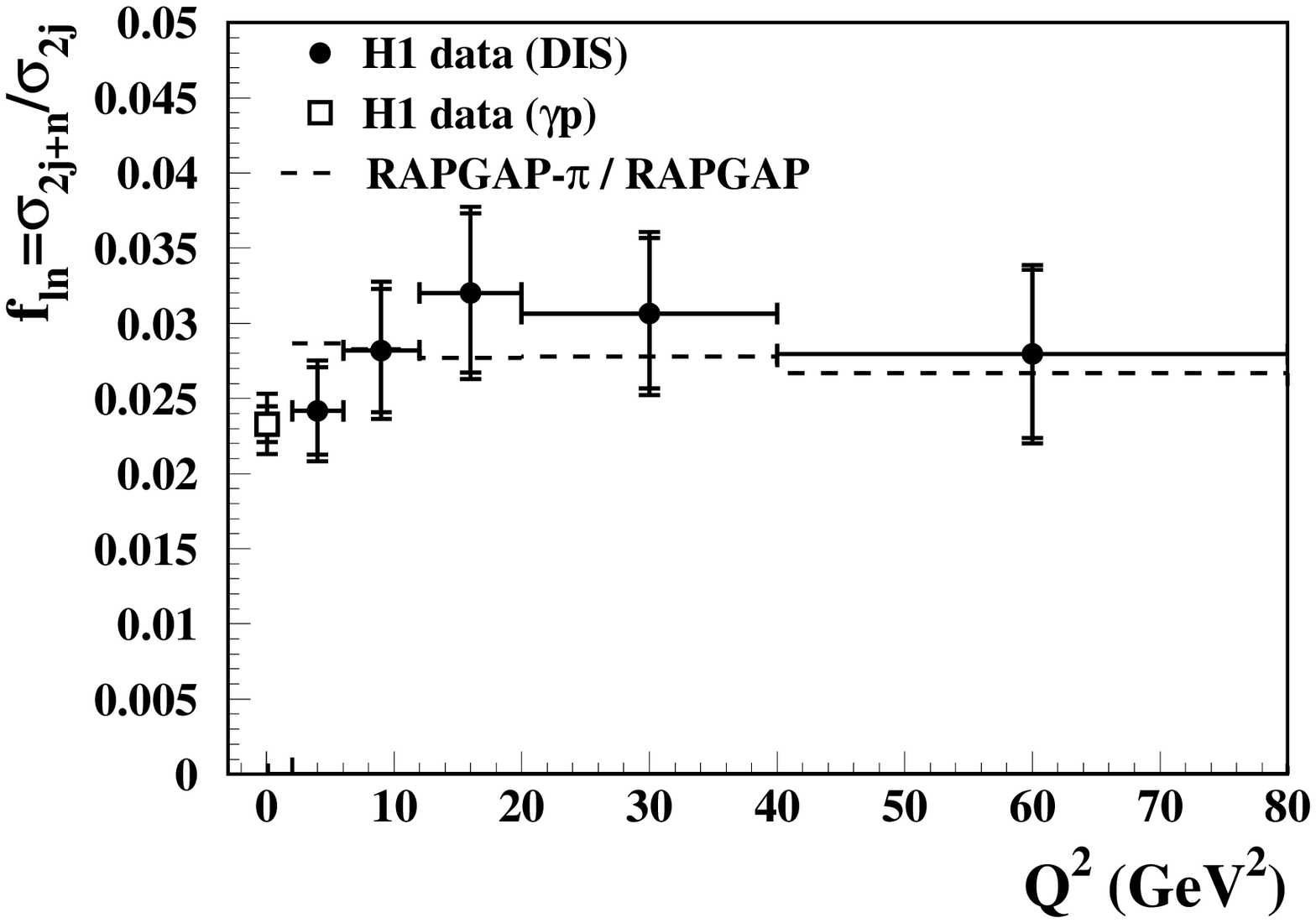,height=80mm}
\caption{The ratio of the cross section for dijet production 
with a leading neutron to that for inclusive dijet production,
as a function of $Q^2$. 
% The open square corresponds to photoproduction.
Inner error bars show the statistical errors, while
the outer error bars represent the statistical and systematic errors,
added in quadrature.
The overall normalization uncertainty of 13\% is not shown.
The Monte Carlo prediction, shown only for DIS, is obtained by using
RAPGAP-$\pi$ for the leading neutron cross section
and RAPGAP for the inclusive cross section.
  The kinematic regions within which this measurement is made
  are given in Table 1.}
\label{ratioq2} 
\end{figure}  

%**************************************************************
The ratio of the dijet cross sections with and without the
requirement of a leading neutron, $f_{ln}$, is an interesting quantity in 
that it discriminates between the various Monte Carlo models used to
describe leading neutron production. Further, if the hard interaction is 
independent of the neutron production, it should be essentially
independent of the jet kinematics which reflect the hard process, neglecting
possible phase space effects.
As the only difference in the event selection for the leading neutron data
and the inclusive dijet samples is the requirement of a leading  neutron, 
some important systematic uncertainties cancel in the ratio. 
The remaining overall normalization uncertainties, mainly associated
with the acceptance, efficiency and energy scale of the FNC calorimeter,
are about 13\%.

For the photoproduction data, $f_{ln}$ is shown in Fig.~\ref{ratio1}
and in Table~4  as a function of  the jet variables.
% $E_T^{jet}$, $\eta_{lab}^{jet}$, $x_\gamma^{jet}$ and  $x_p^{jet}$.  
Figure \ref{ratio1}a shows  that $f_{ln}$ is, within errors, 
independent of $E_T^{jet}$ and has an average value of about 2.3\%.
However, as can be seen in Figs.~\ref{ratio1}b--d, $f_{ln}$
shows a dependence 
on $\eta_{lab}^{jet}$, $x_\gamma^{jet}$ and  $x_p^{jet}$.
%\footnote{The
%quantity $x_p^{jet}$ is defined as 
%$x_p^{jet}=\frac{(E_T^{jet1}e^{\eta^{jet2}}+E_T^{jet2}e^{\eta^{jet2}})}{2E_p}$}.
These dependences can only partly be reproduced by 
the PYTHIA model, which 
provides some estimate of the size of possible phase space effects. 
A better description of the ratio in Fig.~\ref{ratio1}
is possible, if the leading neutron data are described by 
the $\pi$-exchange model, RAPGAP-$\pi$, and the inclusive 
dijet data by PYTHIA-MI. 
This comparison suggests that the mechanism for dijet production in 
events with a leading neutron differs from that in inclusive dijet events.

This is further studied by measuring the
$Q^2$ dependence of the ratio of the DIS dijet cross sections with and 
without the leading neutron requirement.
The result is shown in Fig.~\ref{ratioq2} and in Table~5. 
Here, the point at
$Q^2=0$  is the average of the  ratios for photoproduction
shown in Fig.~\ref{ratio1}.
Within the experimental errors, the RAPGAP model
 describes the measured ratio, when
 the  leading neutron and the inclusive dijet data are
 represented by RAPGAP-$\pi$ and standard RAPGAP, respectively.
 However, there is some tendency for the measured ratio to increase
 with $Q^2$, for $Q^2$ below $20~\GeV^2$.
 A similar $Q^2$ dependence  was observed by the ZEUS Collaboration 
 in the analysis of inclusive DIS events with leading neutrons~\cite{zeuslninc}.

%-----------------------------------------
\subsection{Discussion}

It was observed in \cite{H1LN} that  pion exchange  provides 
a good description of the semi-inclusive DIS process (1) 
in which a leading neutron is produced. The present results, given in 
sections 6.1 to 6.3, demonstrate that this is also the case 
for the small subsample of leading neutron events 
in which a dijet system is produced, in both DIS and photoproduction.
This observation is not trivial, as the parameters used in the Monte Carlo
models to empirically describe  the pion exchange were 
determined in hadronic reactions and no tuning to the present data
was performed.

It is also observed that, for the DIS sample, the standard Monte Carlo 
models for the  simulation of the hadronic final state, such as  LEPTO and 
RAPGAP, predict cross sections for the production of 
dijets with a leading neutron which are too small. 
Again, no attempt to tune parameters 
was made.  The increase of the leading neutron 
rate in dijet production caused by 
the introduction of non-perturbative soft colour interactions in LEPTO-SCI 
is not large enough to provide a good description of the 
measurements. However, LEPTO-SCI successfully describes 
the DIS reaction $ep\rightarrow enX$ \cite{H1LN}.

For the photoproduction sample, equally good descriptions
are obtained with the pion exchange model RAPGAP-$\pi$ 
and the standard Monte Carlo program PYTHIA. 
The predictions of PYTHIA-MI clearly fail to describe the 
leading neutron data. 
However, the introduction of multiple interactions in PYTHIA
is necessary to describe inclusive jet production~\cite{inclusivejet}. 
The $x_\gamma^{jet}$ distributions in Fig.~\ref{xgamma}a demonstrate that 
this discrepancy between the data and PYTHIA-MI predictions is due to the
poor description of resolved photon processes.
The relative fraction of these processes is considerably lower 
in dijet  production with a leading neutron than in inclusive dijet production.
This finding is corroborated by the ratios of the dijet cross sections with
 and without the leading neutron requirement, 
presented in Fig.~\ref{ratio1}c as a function of $x_\gamma^{jet}$.
This ratio increases by a factor of 3 as $x_\gamma^{jet}$ 
increases from 0.3 to 0.9.
The tendency seen in Fig.~\ref{ratioq2} for this ratio to 
rise with $Q^2$
may be due to absorptive corrections, as has been pointed out by several 
authors~\cite{absorption}.

The differences between the dijet production data with and without 
leading neutrons,
as well as the kinematic dependences of the cross 
section ratios, shown in  
Figs.~\ref{ratio1}b--d, 
point to differences in the production mechanism of events with and without
leading neutrons.
The present analysis shows that pion exchange 
is able to describe the properties of leading neutron events.

A similar analysis in the photoproduction regime has been published by 
the ZEUS Collaboration \cite{ZEUSln}. 
Qualitatively, there is good agreement between the two studies. 
A detailed comparison is difficult, however, since the kinematic 
ranges and the jet algorithm used are different.   
The ZEUS Collaboration also observes  that the
ratio of leading neutron events to the inclusive sample
is independent of $E_T^{jet}$, and sees
a similar dependence  of this ratio on $x_\gamma^{jet}$.
The dependence observed in~\cite{ZEUSln}
as a function of $\eta^{jet}$ is somewhat weaker than that in 
Fig.~\ref{ratio1}b.

%%%%%%%%%%%%%%%%%%%%%%%%%%%%%%%%%%%%%%%%%%%%%%%%%%%%%%%%%%%%%%%%%%%%
\section{Summary}

The production of dijet events with a leading neutron
of energy $E_n>500~\GeV$ and polar angle $\theta_n<0.8$~mrad
is studied in photoproduction ($Q^2<10^{-2}~\GeV^2$ and
$0.3<y<0.65$) and in deep inelastic scattering ($2<Q^2<80~\GeV^2$ 
and $0.1<y<0.7$).
Dijet events with $E_T^{jet1}>7~\GeV$ and $E_T^{jet2}>6~\GeV$ 
are selected using a cone algorithm in the $\gamma^*p$  frame.  
The laboratory pseudorapidities of the jets are restricted to the
region  $-1<\eta_{lab}^{jet1,2}<2$.  
Differential cross sections are presented as a function of $E_T^{jet}$, 
$\eta_{lab}^{jet}$, $x_\gamma^{jet}$ and $x_\pi^{jet}$
for photoproduction, and as a function of $Q^2$, $E_T^{jet}$, 
$\eta_{lab}^{jet}$, $x_\gamma^{jet}$ and $x_\pi^{jet}$ for DIS.

Both the cross section measurements and the neutron energy spectrum
are reasonably well described by pion exchange models in which the photon
interacts  with a parton from the exchanged pion.
These models are based on leading order QCD.
The phenomenological parameters describing the pion
exchange are taken from previous analyses of hadronic interactions.
Next-to-leading order QCD calculations, 
after corrections for hadronization effects, 
also describe the measured photoproduction dijet distributions,
in normalization as well as in shape.
The experimental
uncertainties are still too large to discriminate in this kinematic
region  between different parameterizations of the pion parton densities.

Monte Carlo programs which are not based on the pion exchange mechanism,
such as PYTHIA, are also able to describe the leading neutron dijet 
photoproduction data.
However, the predictions of PYTHIA including multiple interactions
fail to describe the measurements.
For the DIS sample, the standard LO Monte Carlo models RAPGAP and
LEPTO, which do not include meson exchange, give a poor 
description of the data. 
This remains true when soft colour interactions are
added to LEPTO, in which case, however, the inclusive leading
neutron data are described.

The ratios of the cross sections with and without the leading neutron
requirement are studied as a function of $Q^2$ and of the jet 
kinematic variables listed above.
If the hard process is independent of the leading neutron
production, these ratios should not depend either on $Q^2$ or 
the jet variables.
Indeed, there is no evidence for a strong dependence of the
ratio on $Q^2$, or on $E_T^{jet}$ in photoproduction.
However, the ratio in photoproduction rises with $x_\gamma^{jet}$.
This suggests that the leading neutron dijet data have a lower
fraction of resolved photon processes
than do the inclusive dijet data.

\section*{Acknowledgements}

We are grateful to the HERA machine group whose outstanding
efforts have made this experiment possible.
We thank the engineers and technicians for their work in constructing and
maintaining the H1 detector, our funding agencies for
financial support, the DESY technical staff for continual assistance
and the DESY directorate for support and for the
hospitality which they extend to the non-DESY members of the 
collaboration.
We wish to thank M.~Klasen and G.~Kramer for making their theoretical
calculations available to us.

%%%%%%%%%%%%%%%%%%%%%%%%%%%%%%%%%%%%%%%%%%%%%%%%%%%%%%%%%%%%%%%

%%%%%%%%%%%%%%%%%%%%%%%%%%%%%%%%%%%%%%%%%%%%%%%%%%%%%%%%%%%%%%%
\newpage
% gp-Et cross section
\begin{table}[h]
\centering
 \label{table1}
  \begin{tabular}{|r r c | c | c | c | c |}
     \hline &   & & & & & \\[-0.4cm]
     \multicolumn{3}{|c|}{Jet Transverse Energy ($E_T^{jet}$)}
            & 
     $d\sigma_{ep}/dE_T^{jet}$ & 
     $\delta_{stat.}$ & 
     $\delta_{uncorrel.syst.}$ & 
     $\delta_{correl.syst.}$ \\
      \multicolumn{3}{|c|}{[GeV]} & [pb/GeV] & [pb/GeV] & [pb/GeV] & [pb/GeV] 
          \\ \hline
    $\qquad$ 6   &  $\qquad$ -  & 8   & 125.  & 7.  & 15.  & 25. \\
    8   &   -  & 10  & 75.4   & 5.2  &  9.1  & 15.1 \\
   10   &   -  & 12  & 37.1   & 3.7  & 4.5  &  7.4 \\   
   12   &   -  & 14  & 13.6   & 2.0  & 1.8  &  2.7 \\   
   14   &   -  & 16  & 4.69   & 1.21 & 0.65 &  0.94 \\   
   16   &   -  & 18  & 2.52   & 0.84  & 0.37  &  0.51 \\   
   18   &   -  & 20  & 2.63   & 0.88  & 0.38  &  0.53 \\   
   20   &   -  & 22  & 2.04   & 0.83  & 0.30  &  0.41 \\   
  \hline 
  \hline &   & & & & & \\[-0.4cm]
     \multicolumn{3}{|c|}{Jet Pseudorapidity ($\eta_{lab}^{jet}$)}
            & 
     $d\sigma_{ep}/d\eta_{lab}^{jet}$ & 
     $\delta_{stat.}$ & 
     $\delta_{uncorrel.syst.}$ & 
     $\delta_{correl.syst.}$ \\
      \multicolumn{3}{|c|}{  } & [pb] & [pb] & [pb] & [pb] 
          \\ \hline
     --1.0 &  -  & --0.5   & 162.  & 18.  & 19.  & 32. \\
     --0.5 &  -  &   0.0   & 190.  & 17.  & 23.  & 38. \\
       0.0 &  -  &   0.5   & 191.  & 17.  & 23.  & 38. \\
       0.5 &  -  &   1.0   & 181.  & 15.  & 22.  & 36. \\
       1.0 &  -  &   1.5   & 180.  & 16.  & 22.  & 36. \\
       1.5 &  -  &   2.0   & 145.  & 13.  & 17.  & 29. \\
  \hline 
  \hline &   & & & & & \\[-0.4cm]
     \multicolumn{3}{|c|}{$x_\gamma^{jet}$}
            & 
     $d\sigma_{ep}/dx_\gamma^{jet}$ & 
     $\delta_{stat.}$ & 
     $\delta_{uncorrel.syst.}$ & 
     $\delta_{correl.syst.}$ \\
      \multicolumn{3}{|c|}{  } & [pb] & [pb] & [pb] & [pb] 
          \\ \hline
       0.2 &  -  &   0.4   & 228.  & 27.  & 31.  & 46. \\
       0.4 &  -  &   0.6   & 249.  & 26.  & 35.  & 50. \\
       0.6 &  -  &   0.8   & 372.  & 40.  & 45.  & 74. \\
       0.8 &  -  &   1.0   & 288.  & 41.  & 35.  & 58. \\
  \hline 
  \hline &   & & & & & \\[-0.4cm]
     \multicolumn{3}{|c|}{$log_{10}(x_\pi^{jet})$}
            & 
     $d\sigma_{ep}/dlog_{10}(x_\pi^{jet})$ & 
     $\delta_{stat.}$ & 
     $\delta_{uncorrel.syst.}$ & 
     $\delta_{correl.syst.}$ \\
      \multicolumn{3}{|c|}{  } & [pb] & [pb] & [pb] & [pb] 
          \\ \hline
       --2.00 &  -  &   --1.67   & 26.6   & 13.3 &  4.0  &  5.3 \\
       --1.67 &  -  &   --1.33   & 171.   & 25.  & 21.  & 34. \\
       --1.33 &  -  &   --1.00   & 281.   & 28.  & 34.  & 56. \\
       --1.00 &  -  &   --0.67   & 239.   & 22.  & 29.  & 48. \\
       --0.67 &  -  &   --0.33   &  38.1  & 5.6  &  5.7  &  7.6 \\
  \hline 
  \end{tabular}
 \caption{The differential $ep$ photoproduction cross section
  as a function of $E_T^{jet}$, $\eta_{lab}^{jet}$, $x_\gamma^{jet}$ and 
  $x_\pi^{jet}$ for dijet events with a leading neutron.
  The kinematic regions within which this measurement is made
  are given in Table 1.}
\end{table}

%%%%%%%%%%%%%%%%%%%%%%%%%%%%%%%%%%%%%%%%%%%%%%%%%%%%%%%
% 13.4% uncorrelated, 20% correlated
% DIS-Et cross section
\begin{table}[h]
\centering
 \label{table5}
  \begin{tabular}{|r r c | c | c | c | c |}
     \hline &   & & & & & \\[-0.4cm]
     \multicolumn{3}{|c|}{Jet Transverse Energy ($E_T^{jet}$)}
            & 
     $d\sigma_{ep}/dE_T^{jet}$ & 
     $\delta_{stat.}$ & 
     $\delta_{uncorrel.syst.}$ & 
     $\delta_{correl.syst.}$ \\
      \multicolumn{3}{|c|}{[GeV]} & [pb/GeV] & [pb/GeV] & [pb/GeV] & [pb/GeV] 
          \\ \hline
    $\qquad$ 6   &  $\qquad$ -  & 8   & 32.2  & 2.5  & 4.3  & 6.4 \\
    8   &   -  & 10  & 24.4                   & 2.2  & 3.3  & 4.9 \\
   10   &   -  & 12  & 14.2                   & 1.8  & 1.9  & 2.8 \\   
   12   &   -  & 14  & 7.8                    & 1.3  & 1.1  & 1.6 \\   
   14   &   -  & 16  & 4.53                   & 1.01 & 0.66 & 0.91 \\   
   16   &   -  & 18  & 2.90                   & 0.92 & 0.43 & 0.58 \\   
  \hline 
  \hline &   & & & & & \\[-0.4cm]
     \multicolumn{3}{|c|}{Jet Pseudorapidity ($\eta_{lab}^{jet}$)}
            & 
     $d\sigma_{ep}/d\eta_{lab}^{jet}$ & 
     $\delta_{stat.}$ & 
     $\delta_{uncorrel.syst.}$ & 
     $\delta_{correl.syst.}$ \\
      \multicolumn{3}{|c|}{  } & [pb] & [pb] & [pb] & [pb] 
          \\ \hline
     --1   &  -  & --0.5   & 45.3  & 7.8  &  6.1  & 9.1 \\
     --0.5 &  -  &   0.0   & 58.1  & 7.2  &  7.8  & 11.6 \\
       0.0 &  -  &   0.5   & 83.2  & 8.5  & 11.1  & 16.6 \\
       0.5 &  -  &   1.0   & 69.7  & 7.3  &  9.3  & 13.9 \\
       1.0 &  -  &   1.5   & 53.9  & 5.8  &  7.2  & 10.8 \\
       1.5 &  -  &   2.0   & 33.1  & 4.5  &  4.4  &  6.6 \\
  \hline 
  \hline &   & & & & & \\[-0.4cm]
     \multicolumn{3}{|c|}{$x_\gamma^{jet}$}
            & 
     $d\sigma_{ep}/dx_\gamma^{jet}$ & 
     $\delta_{stat.}$ & 
     $\delta_{uncorrel.syst.}$ & 
     $\delta_{correl.syst.}$ \\
      \multicolumn{3}{|c|}{  } & [pb] & [pb] & [pb] & [pb] 
          \\ \hline
       0.2 &  -  &   0.4   & 35.0   & 10.1  &  5.3  &  7.0 \\
       0.4 &  -  &   0.6   & 40.8   &  8.6  &  6.2  &  8.2 \\
       0.6 &  -  &   0.8   & 161.  & 24.  & 22.  & 32.  \\
       0.8 &  -  &   1.0   & 190.  & 18.  & 25.  & 38. \\
  \hline 
  \hline &   & & & & & \\[-0.4cm]
     \multicolumn{3}{|c|}{$log_{10}(x_\pi^{jet})$}
            & 
     $d\sigma_{ep}/dlog_{10}(x_\pi^{jet})$ & 
     $\delta_{stat.}$ & 
     $\delta_{uncorrel.syst.}$ & 
     $\delta_{correl.syst.}$ \\
      \multicolumn{3}{|c|}{  } & [pb] & [pb] & [pb] & [pb] 
          \\ \hline
       --2.00 &  -  &   --1.67   &  5.0  &  3.5  &  0.8  &  1.0 \\
       --1.67 &  -  &   --1.33   & 56.   & 12.   &  8.   & 11.  \\
       --1.33 &  -  &   --1.00   & 82.   & 11.   & 11.   & 16.  \\
       --1.00 &  -  &   --0.67   & 84.   & 10.   & 12.   & 17.  \\
       --0.67 &  -  &   --0.33   & 15.8  &  3.2  &  2.4  &  3.2 \\
  \hline 
  \hline &   & & & & & \\[-0.4cm]
     \multicolumn{3}{|c|}{$Q^2$}
            & 
     $d\sigma_{ep}/dQ^2$ & 
     $\delta_{stat.}$ & 
     $\delta_{uncorrel.syst.}$ & 
     $\delta_{correl.syst.}$ \\
      \multicolumn{3}{|c|}{[GeV$^2$]} & [pb/GeV$^2$] & [pb/GeV$^2$] & [pb/GeV$^2$] & [pb/GeV$^2$] 
          \\ \hline
    2   &   -  & 6   & 7.06    & 0.88  & 0.94  & 1.40 \\
    6   &   -  & 12  & 2.82    & 0.41  & 0.38  & 0.56 \\
   12   &   -  & 20  & 1.51    & 0.25  & 0.20  & 0.30 \\   
   20   &   -  & 30  & 0.95    & 0.18  & 0.13  & 0.19  \\   
   30   &   -  & 40  & 0.408   & 0.123 & 0.059 & 0.082 \\   
   40   &   -  & 50  & 0.319   & 0.106 & 0.046 & 0.064 \\   
   50   &   -  & 60  & 0.277   & 0.098 & 0.042 & 0.055 \\   
   60   &   -  & 80  & 0.153   & 0.054 & 0.023 & 0.031 \\   
  \hline 
  \end{tabular}
 \caption{The differential deep inelastic $ep$ cross section
  as a function of $E_T^{jet}$, $\eta_{lab}^{jet}$, $x_\gamma^{jet}$,
  $x_\pi^{jet}$ and $Q^2$ for dijet events with a leading neutron.
  The kinematic regions within which this measurement is made
  are given in Table 1.}
\end{table}

%-----------------------------------------------------------------------
% gp Et, ratio
% uncorr (11.2-15.8)  correlated 12.7%
\begin{table}[h]
\centering
 \label{table11}
  \begin{tabular}{|r r c | c | c | c | c |}
     \hline &   & & & & & \\[-0.4cm]
     \multicolumn{3}{|c|}{Jet Transverse Energy ($E_T^{jet}$)}
            & 
     $f_{ln}$ & 
     $\delta_{stat.}$ & 
     $\delta_{uncorrel.syst.}$ & 
     $\delta_{correl.syst.}$ \\
      \multicolumn{3}{|c|}{[GeV]} & & & & \\ \hline
    $\qquad$ 6   &  $\qquad$ -  & 8   & 0.0231  & 0.0012  & 0.0026  & 0.0030 \\
    8   &   -  & 10  & 0.0244   & 0.0017  & 0.0027  & 0.0031 \\
   10   &   -  & 12  & 0.0200   & 0.0020  & 0.0032  & 0.0025 \\
   12   &   -  & 14  & 0.0202   & 0.0030  & 0.0032  & 0.0026 \\
   14   &   -  & 16  & 0.0212   & 0.0055  & 0.0033  & 0.0027 \\
   16   &   -  & 18  & 0.0152   & 0.0050  & 0.0024  & 0.0019 \\
  \hline 
  \hline &   & & & & & \\[-0.4cm]
     \multicolumn{3}{|c|}{Jet Pseudorapidity ($\eta_{lab}^{jet}$)}
            & 
     $f_{ln}$ &  
     $\delta_{stat.}$ & 
     $\delta_{uncorrel.syst.}$ & 
     $\delta_{correl.syst.}$ \\ &   & & & & & \\[-0.4cm]
\hline
     --1.0 &  -  & --0.5   & 0.0363   & 0.0041  & 0.0041  & 0.0046  \\
     --0.5 &  -  &   0.0   & 0.0277   & 0.0025  & 0.0031  & 0.0035  \\
       0.0 &  -  &   0.5   & 0.0221   & 0.0019  & 0.0027  & 0.0028  \\
       0.5 &  -  &   1.0   & 0.0218   & 0.0018  & 0.0027  & 0.0028  \\
       1.0 &  -  &   1.5   & 0.0211   & 0.0018  & 0.0027  & 0.0027  \\
       1.5 &  -  &   2.0   & 0.0143   & 0.0012  & 0.0018  & 0.0018 \\
  \hline 
  \hline &   & & & & & \\[-0.4cm]
     \multicolumn{3}{|c|}{$x_\gamma^{jet}$}
            & 
     $f_{ln}$ & 
     $\delta_{stat.}$ & 
     $\delta_{uncorrel.syst.}$ & 
     $\delta_{correl.syst.}$ \\ &   & & & & & \\[-0.4cm]
\hline
       0.2 &  -  &   0.4   & 0.0105   & 0.0013  & 0.0013  & 0.0013 \\
       0.4 &  -  &   0.6   & 0.0256   & 0.0027  & 0.0029  & 0.0033 \\
       0.6 &  -  &   0.8   & 0.0340   & 0.0037  & 0.0038  & 0.0044 \\
       0.8 &  -  &   1.0   & 0.0344   & 0.0049  & 0.0039  & 0.0045 \\
  \hline 
  \hline &   & & & & & \\[-0.4cm]
     \multicolumn{3}{|c|}{$log_{10}(x_p^{jet})$}
            & 
     $f_{ln}$ & 
     $\delta_{stat.}$ & 
     $\delta_{uncorrel.syst.}$ & 
     $\delta_{correl.syst.}$ \\  &   & & & & & \\[-0.4cm]
\hline
       --2.33  &  - &   --2.00  & 0.0367  & 0.0057  &  0.0043  & 0.0047 \\
       --2.00  & -  &   --1.67  & 0.0216  & 0.0019  &  0.0024  & 0.0027 \\
       --1.67 &  -  &   --1.33  & 0.0208  & 0.0016  &  0.0023  & 0.0026 \\
       --1.33 &  -  &   --1.00  & 0.0133  & 0.0021  &  0.0015  & 0.0017 \\
  \hline 
  \end{tabular}
 \caption{The ratio of the cross section for dijet photoproduction
  with a leading neutron to that for inclusive dijet photoproduction,
  as a function of
  $E_T^{jet}$, $\eta_{lab}^{jet}$, $x_{\gamma}^{jet}$  and $x_p^{jet}$.
  The kinematic regions within which this measurement is made
  are given in Table 1.}

\end{table}

%*************************************************************
% DIS-Q2 ratio
% for DIS uncorrelated 7%  , 12.7 correlated
\begin{table}[h]
\centering
 \label{table14}
  \begin{tabular}{|c c c | c | c | c | c |}
     \hline &   & & & & & \\[-0.4cm]
     \multicolumn{3}{|c|}{$Q^2$} &  $f_{ln}$ & $\delta_{stat.}$ & 
     $\delta_{uncorrel.syst.}$ &   $\delta_{correl.syst.}$ \\       
      \multicolumn{3}{|c|}{[GeV$^2$]} & & & &  \\ \hline &   & & & & & \\[-0.4cm]
      \multicolumn{3}{|c|}{$<10^{-2}$} & 0.0233 & 0.0012 & 0.0016 & 0.0029  \\ 
    ~~~~~2~~~~~   &   ~~~~~-~~~~~  & ~~~~~6~~~~~   & 
                        0.0241   & 0.0029  & 0.0017  & 0.0031 \\
    6   &   -  & 12  & 0.0282   & 0.0041  & 0.0020  & 0.0061 \\
   12   &   -  & 20  & 0.0320   & 0.0053  & 0.0022  & 0.0041 \\
   20   &   -  & 40  & 0.0307   & 0.0050  & 0.0021  & 0.0039 \\
   40   &   -  & 80  & 0.0279   & 0.0056  & 0.0020  & 0.0036 \\
  \hline 
  \end{tabular}
 \caption{The ratio of the cross section for dijet production
  with a leading neutron to that for inclusive dijet production,
  as a function of $Q^2$.
  The kinematic regions within which this measurement is made
  are given in Table 1.}
\end{table}

\end{document}